\documentclass[a4paper,11pt]{article}
\usepackage{jheppub}
\usepackage{soul}
\usepackage{comment}
\usepackage{orcidlink}
\usepackage{mdframed}
\usepackage[normalem]{ulem}
\allowdisplaybreaks[3]

\preprint{{\small USTC-ICTS/PCFT-26-29}}

\makeatletter
\def\@fpheader{~}
\makeatother

\title{\boldmath Quantum-Corrected Q-balls in the Friedberg-Lee-Sirlin Model}

\author[a]{Yong-Xiang Su,}
\author[b]{Qi-Xin Xie \orcidlink{0009-0000-2391-863X},}
\author[c,d]{and Shuang-Yong Zhou}
\affiliation[a]{School of Physical Sciences, University of Science and Technology of China, Hefei, Anhui 230026, China}
\affiliation[b]{School of Physics and Astronomy, University Park, University of Nottingham,\\ Nottingham NG7 2RD, United Kingdom}
\affiliation[c]{Interdisciplinary Center for Theoretical Study, University of
Science and Technology of China, Hefei, Anhui 230026, China}
\affiliation[d]{Peng Huanwu Center for Fundamental Theory, Hefei, Anhui 230026, China}

\emailAdd{anonym@mail.ustc.edu.cn}
\emailAdd{qixin.xie@nottingham.ac.uk}
\emailAdd{zhoushy@ustc.edu.cn}

\abstract{We study the real-time quantum dynamics of Q-balls in the Friedberg-Lee-Sirlin model
within the inhomogeneous Hartree approximation. The mean fields are evolved
self-consistently with the leading quantum two-point functions, which are implemented
numerically through a stochastic ensemble representation. After introducing a
renormalized formulation and a classical-limit scaling, we simulate single-Q-ball
configurations in $3+1$ dimensions and compare their quantum-corrected evolution with
the corresponding classical dynamics. We find a clear separation between a classical regime,
where quantum fluctuations remain small and the evolution closely follows the classical
solution, and a quantum regime, where the fluctuation sector carries a sizable fraction
of the Noether charge. We also observe a periodic exchange of Noether charge between the mean fields and the fluctuation modes within the Hartree approximation. We further investigate the stability of quantum-corrected Q-balls and
find an intermediate window in which configurations that are classically stable become
unstable once Hartree fluctuations are included. Our results provide a first step toward
real-time quantum simulations of Q-balls in renormalizable two-field soliton models.}

\begin{document}
\maketitle
\flushbottom

\section{Introduction}
\label{sec:intro}

Non-topological solitons, in particular Q-balls \cite{Friedberg:1976me, Coleman:1985ki, Rosen:1968mfz} (see \cite{Zhou:2024mea} for a recent review), are spatially localized field configurations that can persist for long times in nonlinear field theories, even without topological protection. Their stability is instead associated with conserved Noether charges arising from continuous symmetries. They exhibit rich dynamics, especially in multi-soliton systems \cite{Axenides:1999hs,Battye:2000qj,Bowcock:2008dn,Blaschke:2024dlt,Martinez:2025ana}. Depending on their relative phases and the signs and magnitudes of their charges, they can attract, repel, or exchange charge with one another \cite{Axenides:1999hs,Battye:2000qj}. There also exist rotational solitons \cite{Volkov:2002aj,Kleihaus:2005me} and composite structures such as charge-swapping Q-balls (CSQs) \cite{Copeland:2014qra,Xie:2021glp,Hou:2022jcd}. Non-topological solitons have found many applications in particle physics and cosmology; see, for example, \cite{Friedberg:1977xf,Friedberg:1978sc,Rho:1983bh,Cahill:1985mh,Kusenko:1997si,Enqvist:1997si,Kasuya:1999wu,Kasuya:2000sc,Kusenko:2001vu,Fujii:2001xp,Multamaki:2002hv,Kawasaki:2002hq,Roszkowski:2006kw,Kusenko:2008zm,Hartmann:2012wa,Kasuya:2014bxa,Boskovic:2021nfs,Hou:2022jcd,Pearce:2022ovj,Kawasaki:2023rfx}.

A particularly important class is provided by single-field Q-balls, which can be realized in theories of a single complex scalar when higher-dimensional effective operators are included. Early mathematical constructions of this type can be traced back to Rosen \cite{Rosen:1968mfz} in 1968, while their stability mechanism and physical relevance were later clarified by Coleman \cite{Coleman:1985ki} in 1985. In this work, however, we focus on the Friedberg-Lee-Sirlin (FLS) model, which historically appeared before the single-field Q-ball construction \cite{Friedberg:1976me} and is given by a renormalizable two-scalar theory, containing one real and one complex scalar, with a symmetry-breaking potential. Its non-topological solitons are stabilized by the U(1) charge carried by the complex field. It remains one of the simplest UV-complete settings. Various properties of FLS solitons and their variants have been investigated in \cite{Friedberg:1976az,Friedberg:1976ay,Friedberg:1976eg,Lee:1986tr,Lee:1991ax,Lensky:2001xy,Gani:2004oyh, Levin:2010gp, Loiko:2018mhb, Heeck:2023idx, Azatov:2024npx}. For instance, gauged FLS solitons also exist when the complex scalar is coupled to electromagnetism \cite{Lee:1991bn}, and axially symmetric configurations in such models can carry both electric and magnetic fields \cite{Loiko:2019gwk}. When gravity is included, the corresponding objects are boson stars, which have also been constructed and studied in the FLS model \cite{Friedberg:1986tq,Kunz:2019sgn,Kunz:2021mbm,Kunz:2023qfg,deSa:2024dhj,Jaramillo:2024cus}.

Most studies of Q-balls have focused on classical dynamics, but their quantum properties have also been investigated \cite{Friedberg:1976me,Kusenko:1997ad,Graham:2001hr,Tranberg:2013cka,Levkov:2017paj,Kovtun:2018jae,Kovtun:2021rcm,Xie:2023psz,Ogundipe:2024chv,Kim:2024vam, Evslin:2025hjt}. In addition, many existing analyses focus either on the near-classical regime \cite{Friedberg:1976me,Kusenko:1997ad}, static one-loop quantities \cite{Graham:2001hr,Kim:2024vam}, or large-$N$ approximations \cite{Kovtun:2021rcm}. In contrast, real-time lattice simulations can be used to probe the highly nonperturbative dynamics of quantum Q-balls. 

The inhomogeneous Hartree approximation \cite{Salle:2000hd} offers a practical framework for this purpose. In this approach, the classical field is promoted to the mean quantum field, while connected correlation functions are retained up to the level of two-point functions. The mean fields and two-point functions then evolve self-consistently. Equivalently, the same equations can be derived as the leading-order truncation of the two-particle-irreducible (2PI) effective action \cite{Cornwall:1974vz,Berges:2004yj}. This method has been successfully applied to a variety of inhomogeneous systems \cite{Bettencourt:2001id,Bettencourt:2001xg,Bergner:2002we,Bergner:2003au,Salle:2003ju,Borsanyi:2007wm,Saffin:2014yka,Tranberg:2013cka,Xie:2023psz}. It can also be combined with the ensemble method \cite{Borsanyi:2007wm,Borsanyi:2008eu,Berges:2010zv,Saffin:2011kc,Saffin:2011kn,Hebenstreit:2013qxa}, in which quantum mode functions are replaced by classical stochastic fields with the same statistical properties. In the large-ensemble limit, this stochastic representation reproduces the Hartree two-point functions.

The Hartree approximation has been used to study the dynamics and stability of single-field Q-balls in theories with a sixth-order complex-scalar potential \cite{Tranberg:2013cka,Xie:2023psz}. In those studies, as expected, large Q-balls behave almost classically, whereas small Q-balls display pronounced quantum effects. Quantum-corrected interactions of multiple Q-balls have also been investigated \cite{Xie:2023psz}. Closely related objects such as oscillons display similar dynamical behavior, and their quantum stability has been studied within the same framework \cite{Saffin:2014yka}; quantum effects in boson stars have also been explored in \cite{Saffin:2026tvg}.

In this paper, we study the real-time quantum dynamics of Q-balls in the FLS model in $3+1$ dimensions using the inhomogeneous Hartree approximation. Our main goal is to understand how quantum fluctuations modify their evolution and stability beyond the classical approximation.  More specifically, we determine when the quantum evolution remains close to the classical evolution, and when genuine quantum effects become important. We also study the stability of configurations with sizable quantum corrections and compare them with the stability criterion from classical solutions. We find a clear separation between classical and quantum regimes, and observe a periodic exchange of Noether charge between the mean fields and the fluctuation modes. We also identify a metastable window in which a classically stable Q-ball becomes unstable once Hartree fluctuations are included.

The paper is organized as follows. In section~\ref{sec:model}, we introduce the FLS model, review the classical Q-ball solutions, and present the inhomogeneous Hartree approximation, the stochastic ensemble method, the renormalization procedure, and the numerical observables used in our simulations. In section~\ref{sec:Results}, we present the numerical results, beginning with classical dynamics and then moving to quantum evolution in both the classical and quantum regimes, followed by a stability analysis. We conclude in section~\ref{sec:summary} with a discussion of the main results and possible future directions. In appendix~\ref{app:2pi}, we derive the Hartree approximation used in the main text from the leading-order truncation of the $2$PI effective action.

\section{Model and Setup}
\label{sec:model}
In this section, we introduce the Friedberg-Lee-Sirlin (FLS) model and describe the inhomogeneous Hartree approximation used to incorporate leading quantum corrections. We then present the numerical implementation and the observables used in the simulations.

\subsection{Friedberg-Lee-Sirlin Model}
\label{sec:FLS}

The FLS model \cite{Friedberg:1976me} contains one complex scalar field $\phi$ and one real scalar field $\chi$. Its Lagrangian is
\begin{equation} 
\label{eq:1}
\mathcal{L} =|\partial _{\mu}\phi |^2+\frac{1}{2}\partial _{\mu}\chi \partial ^{\mu}\chi -U\left( |\phi |,\chi \right) ,
\end{equation}
where the Minkowski metric is $\eta_{\mu\nu}={\rm diag}(+---)$, and we use natural units $\hbar=c=1$. The standard potential that supports soliton solutions is
\begin{equation}
\label{eq:4}
U\left( |\phi |,\chi \right) =h^2\chi ^2|\phi |^2+\frac{1}{8}g^2\left( \chi ^2-\chi _{v}^{2} \right) ^2,
\end{equation}
which spontaneously breaks the $\mathbb{Z}_2$ symmetry of $\chi$ and gives a nonzero vacuum value $\chi_v$. The global $U(1)$ symmetry $\phi \rightarrow e^{\mathrm{i}\theta}\phi$ gives the conserved Noether current 
$j^\mu=\mathrm{i}(\phi\partial^\mu\phi^*-\phi^*\partial^\mu\phi)$. Its temporal component defines the conserved charge
\begin{equation}
\label{eq:13}
Q=\int{\mathrm{d}^3xj^0}.
\end{equation}
Time-translation symmetry gives the conserved energy
\begin{equation}
\label{eq:14}
E=\int{\mathrm{d}^3x \left[|\partial_t\phi|^2+|\nabla\phi|^2+\frac{1}{2}(\partial_t\chi)^2+\frac{1}{2}(\nabla\chi)^2+U(|\phi|,\chi)\right]}.
\end{equation}

The complex scalar can equivalently be written in terms of its real and imaginary components
\begin{equation}
    \label{eq:34}
    \phi =\frac{1}{\sqrt{2}}\left( \phi _1+\mathrm{i}\phi _2 \right) .
\end{equation}
The equations of motion for the three real fields
$\phi_1$, $\phi_2$, and $\chi$ are
\begin{align}
    \label{eq:eom1}
& \left( \partial _{x}^{2}+h^2\chi ^2 \right) \phi _i\left( x \right) =0,~~i=1,2,
\\
    \label{eq:eom3}
& \left[ \partial _{x}^{2}+h^2\left( \phi _{1}^{2}+\phi _{2}^{2} \right) +\frac{1}{2}g^2\left( \chi ^2-\chi _{v}^{2} \right) \right] \chi \left( x \right) =0.
\end{align}
Expanding around the vacuum $\phi_i=0,\chi=\chi_v$, one finds the field masses $m_{\chi}=g\chi _v$ and $m_{\phi}=h\chi _v$. Thus, although the Lagrangian contains no explicit mass term for $\phi$, the vacuum value $\chi_v$ generates an effective mass for the complex field.

The nonlinear interactions in this model support solitonic solutions. We focus on configurations of the form
\begin{equation}
\label{eq:6} \phi \left( \boldsymbol{x},t \right) =\frac{1}{\sqrt{2}}e^{\mathrm{i}\omega t}f\left( r \right), \quad \chi \left( \boldsymbol{x},t \right) =\xi \left( r \right).
\end{equation}
The real functions $f(r)$ and $\xi(r)$ are localized in space. The parameter $\omega$ is the internal rotation frequency of the complex field. For solitons with exponential falloff at spatial infinity, the frequency must be smaller than the field mass $|\omega|<m_{\phi}$. By time-reversal symmetry, we can restrict to $\omega>0$ without loss of generality. Regularity at the origin requires the derivatives of $f(r)$ and $\xi(r)$ to vanish
\begin{equation}
    f'(0)=\xi'(0)=0,
\end{equation}
where $'$ denotes differentiation with respect to $r$. Asymptotic approach to the vacuum requires
\begin{equation}
    f(\infty)=0, \qquad\xi(\infty)=\chi_v.
\end{equation}

A central question for any soliton solution is its stability. In classical dynamics, stability against perturbations is characterized by $\mathrm{d}Q/\mathrm{d}\omega<0$, which gives an upper bound on the frequency $\omega<\omega_c$. Quantum effects further restrict the stable region. To prevent decay into free particles, the energy per charge must satisfy 
\begin{equation}
    E<m_{\phi}Q.
\end{equation}
This condition defines a separate critical frequency $\omega_\phi$ through $E(\omega_\phi)=m_\phi Q(\omega_\phi)$. Stable FLS solitons are minimum-energy configurations at fixed charge.

Compared with single-field Q-balls, which typically require a nonrenormalizable potential for existence, the potential in Eq.~\eqref{eq:4} is power-counting renormalizable because it contains at most four powers of the scalar fields. This renormalizability is achieved by introducing the additional real field $\chi$. Nevertheless, FLS solitons share many qualitative features with polynomial Q-balls \cite{Coleman:1985ki}. For example, the charge diverges near the lower and upper frequency endpoints, and the stability regions are separated by critical frequencies: $\omega_c$ for classical stability and $\omega_{\phi}$ for quantum stability. When the quantum corrections are included, these stability regions persist, but their locations are shifted, as we will show in later sections. 

For the quantum calculation, the restricted potential in Eq.~\eqref{eq:4}
is not closed under renormalization, since the allowed mass counterterms are
generated by ultraviolet divergences. We therefore make a minimal extension by adding explicit mass terms and use instead the potential
\begin{equation}
\label{eq:quantum pot}
U\left( |\phi |,\chi \right) =h^2\chi ^2|\phi |^2+w_{1}^{2}|\phi |^2+\frac{1}{2}w_{2}^{2}\chi ^2+\frac{1}{8}g^2(\chi ^2-\chi _{v}^{2})^2.
\end{equation}
The corresponding equations of motion are
\begin{equation}
\label{eq:35}
    \left( \partial _{x}^{2}+w_{1}^{2}+h^2\chi ^2 \right) \phi _1\left( x \right) =0,
\end{equation}
\begin{equation}
\label{eq:36}
    \left( \partial _{x}^{2}+w_{1}^{2}+h^2\chi ^2 \right) \phi _2\left( x \right) =0,
\end{equation}
\begin{equation}
\label{eq:37}
\left[ \partial _{x}^{2}+w_{2}^{2}+h^2\left( \phi _{1}^{2}+\phi _{2}^{2} \right) +\frac{1}{2}g^2\left( \chi ^2-\chi _{v}^{2} \right) \right] \chi \left( x \right) =0.
\end{equation}
This modification preserves the original symmetries, and the original Q-ball solutions are recovered when $w_1=w_2=0$. In what follows, we use the shorthand $\phi _3=\chi$.

\subsubsection{Classical Soliton Profiles}
\label{sec:classicalSoliton}

Because the classical equations are unchanged when the Lagrangian is multiplied by an overall constant, it is convenient to express all quantities in units of $\chi_v$. In this subsection we therefore use 
\begin{equation}
    x\to x\chi_v,~~\omega\to\frac{\omega}{\chi_v}~~\phi\to\frac{\phi}{\chi_v},~~\chi\to\frac{\chi}{\chi_v},~~\chi_v\to1.
\end{equation}
Other derived quantities can be made dimensionless in the same way; for example, $E\to E/\chi_v$. 

Eqs.~\eqref{eq:eom1} and~\eqref{eq:eom3} are highly nonlinear, so we solve them numerically. To obtain the soliton profiles, a relaxation algorithm is used \cite{press2007numerical}. The fields are represented by discrete lattice functions and the derivatives are replaced by finite differences. Starting from an initial guess, the algorithm iteratively refines the fields until both the absolute and relative errors fall below $10^{-10}$. 

\begin{figure}[tbp]
\centering
\includegraphics[width=.42\textwidth]{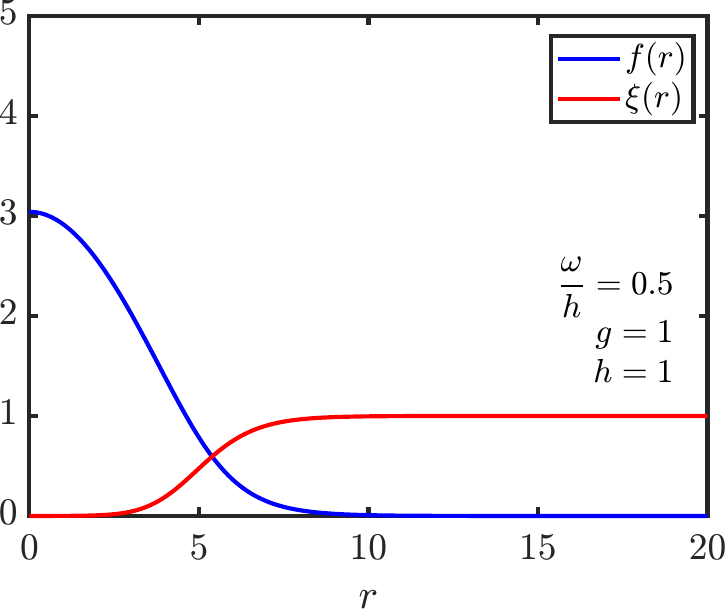}
\hfill
\includegraphics[width=.425\textwidth]{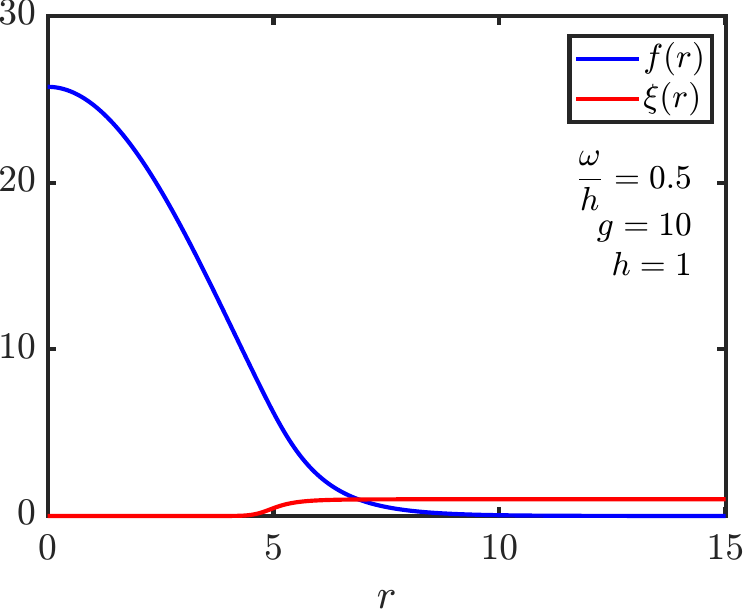}
\hfill
\includegraphics[width=.42\textwidth]{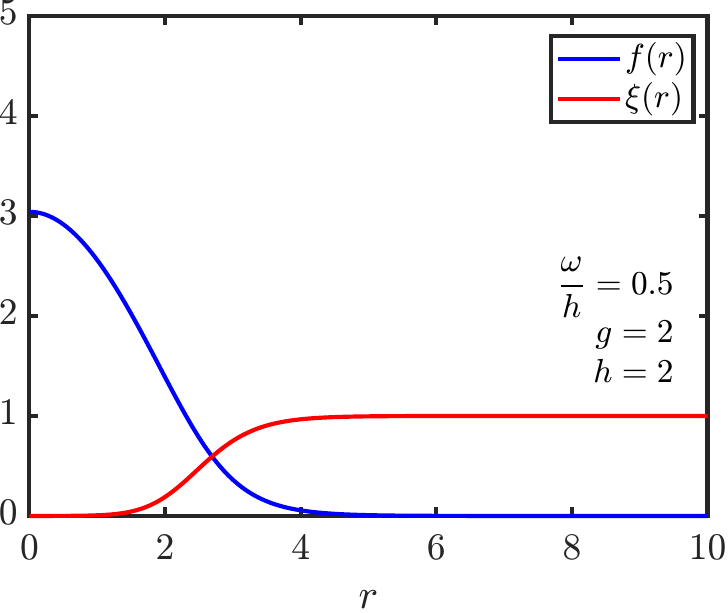}
\hfill
\includegraphics[width=.425\textwidth]{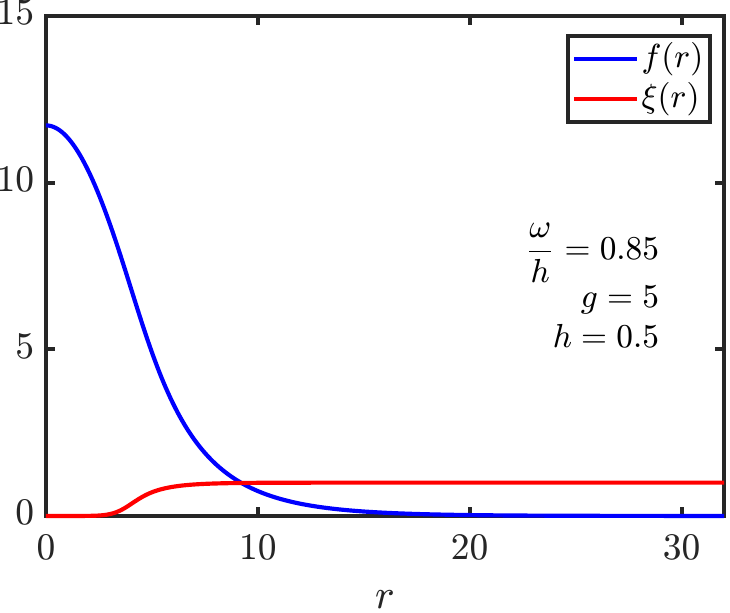}
\caption{Classical soliton profiles for representative parameter choices.  \label{fig:1}}
\end{figure}

The radial profiles $f$ and $\xi$ are shown in figure~\ref{fig:1}. At large distances, both fields approach their vacuum values, while near the center they deviate substantially and produce a localized energy density. As the frequency increases, the Q-ball radius decreases and the central value of the complex scalar profile also becomes smaller. By contrast, increasing $g$ raises the central value of the complex scalar profile. For fixed $h/g$ and $\omega/h$, changing the overall scale simply rescales the radial coordinate $r$, leaving the field amplitudes unchanged. Low-frequency Q-balls have relatively thin surface layers, whereas high-frequency Q-balls have thicker surfaces.

\begin{figure}[tbp]
\centering
\includegraphics[width=.42\textwidth]{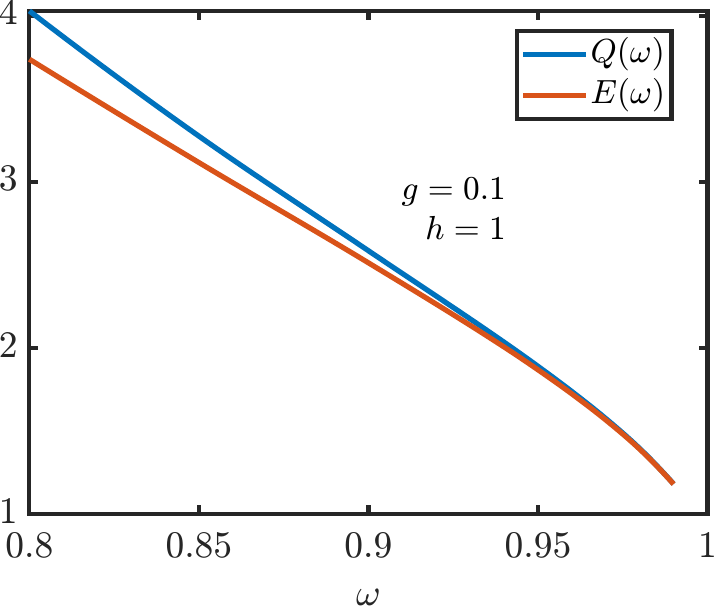}
\hfill
\includegraphics[width=.425\textwidth]{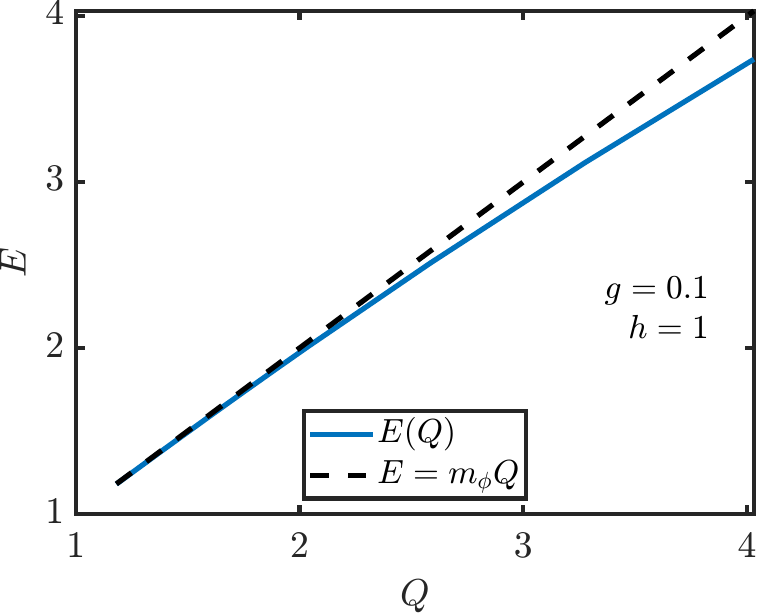}
\hfill
\includegraphics[width=.455\textwidth]{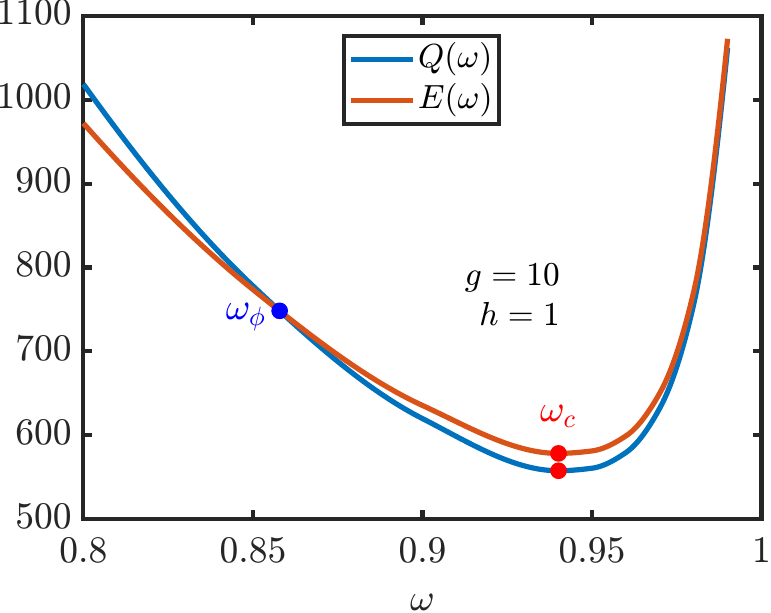}
\hfill
\includegraphics[width=.46\textwidth]{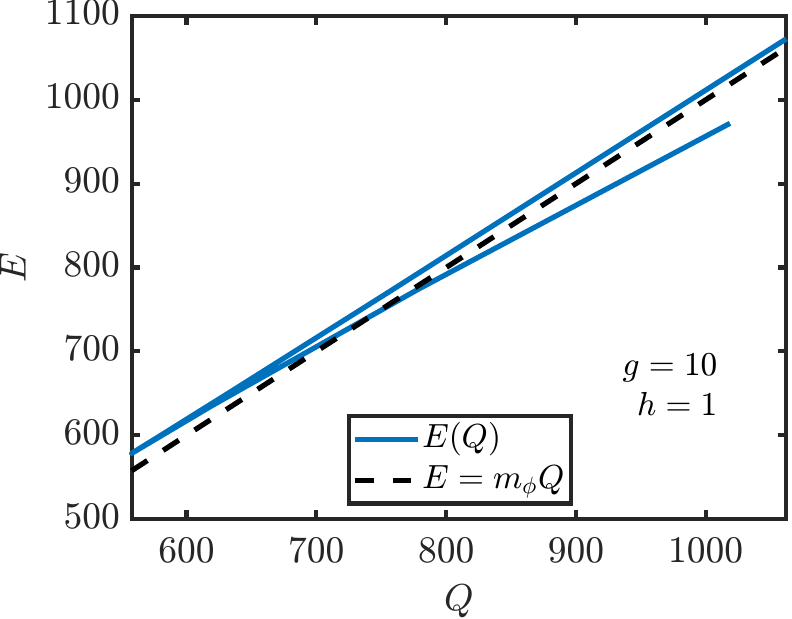}
\caption{Total energy and charge as functions of frequency for representative parameter choices ($E$ and $Q$ are in units of $2\pi$).\label{fig:4} }
\end{figure}

The total charge and energy are shown in figure~\ref{fig:4}. For small $g$ (e.g. $0.1$), both the charge and the energy decrease monotonically with frequency \cite{Loiko:2018mhb,Jaramillo:2024cus}. As $g$ increases, the overall charge and energy also increase. For large $g$, a two-branch structure appears: the charge and energy reach minima at the same critical frequency $\omega_c$ and then rise again as $\omega$ approaches either endpoint. Similarly, for small $g$ the energy increases monotonically with the charge, whereas for larger $g$ the $E(Q)$ curve splits into two branches meeting at a sharp cusp at $\omega_c$. The lower branch corresponds to $\omega<\omega_c$, while the upper branch corresponds to $\omega>\omega_c$. Configurations on the upper branch are unstable under small perturbations. Dashed lines correspond to the energy-charge relation of plane waves, whose intersection point with the Q-ball energy-charge relation at $\omega_{\phi}$ separates the quantum-mechanically stable lower branch from the unstable branch, which decays by emitting free particles. 

\subsection{Inhomogeneous Hartree Approximation}
\label{sec:hatree}
The Hartree approximation is a well-known method to incorporate leading quantum dynamical effects. It was extended in \cite{Salle:2000hd} to inhomogeneous mean fields. The equations of motion of the mean fields and the two-point correlation functions are obtained by dropping connected correlators of order higher than two, thereby closing the equations. Equivalently, the same closed set of equations can be derived from the local double-bubble, or Hartree, truncation of the two-particle-irreducible effective action, in which the mean
fields and the connected two-point functions are treated as independent variables and are
determined by the stationary conditions of the truncated 2PI effective action. This provides a parallel self-consistent variational formulation of the Hartree approximation. We present this derivation in Appendix~\ref{app:2pi}.

The basic variables in the Hartree approximation are the one-point functions, or mean fields
\begin{equation}
    \label{eq:38}
    \Phi _1\left( x \right) =\left< \phi _1\left( x \right) \right> ,  ~~  \Phi _2\left( x \right) =\left< \phi _2\left( x \right) \right> ,  ~~  \Phi _3\left( x \right) =\left< \chi \left( x \right) \right> ,
\end{equation}
together with the connected two-point Wightman functions
\begin{align}
    \label{eq:39}
   G_1( x,y ) &=\left< ( \phi _1( x ) -\Phi _1( x ) ) ( \phi _1( y ) -\Phi _1( y ) ) \right> =\left< \phi _1( x ) \phi _1( y ) \right> -\left< \phi _1( x ) \right> \left< \phi _1( y ) \right> ,
\\
    \label{eq:40}
   G_2\left( x,y \right) &=\left< ( \phi _2( x ) -\Phi _2( x ) ) ( \phi _2( y ) -\Phi _2( y ) ) \right> =\left< \phi _2( x ) \phi _2( y ) \right> -\left< \phi _2( x ) \right> \left< \phi _2( y ) \right> ,
\\
    \label{eq:41}
G_3\left( x,y \right) &=\left< \left( \chi \left( x \right) -\Phi _3\left( x \right) \right) \left( \chi \left( y \right) -\Phi _3\left( y \right) \right) \right> =\left< \chi \left( x \right) \chi \left( y \right) \right> -\left< \chi \left( x \right) \right> \left< \chi \left( y \right) \right> ,
\\
    \label{eq:42}
K_1( x,y ) &=\left< ( \phi _1( x ) -\Phi _1( x ) ) ( \phi _2( y ) -\Phi _2( y ) ) \right> =\left< \phi _1( x ) \phi _2( y ) \right> -\left< \phi _1( x ) \right> \left< \phi _2( y ) \right>, 
\\
\label{eq:43}
\bar{K}_1( x,y ) &=\left< ( \phi _2( x ) -\Phi _2( x ) ) ( \phi _1( y ) -\Phi _1( y ) ) \right> =\left< \phi _2( x ) \phi _1( y ) \right> -\left< \phi _2( x ) \right> \left< \phi _1( y ) \right> ,
\\
    \label{eq:44}
K_2\left( x,y \right) &=\left< \left( \phi _1\left( x \right) -\Phi _1\left( x \right) \right) \left( \chi \left( y \right) -\Phi _3\left( y \right) \right) \right> =\left< \phi _1\left( x \right) \chi \left( y \right) \right> -\left< \phi _1\left( x \right) \right> \left< \chi \left( y \right) \right> ,
\\
    \label{eq:45}
\bar{K}_2\left( x,y \right) &=\left< \left( \chi \left( x \right) -\Phi _3\left( x \right) \right) \left( \phi _1\left( y \right) -\Phi _1\left( y \right) \right) \right> =\left< \chi \left( x \right) \phi _1\left( y \right) \right> -\left< \chi \left( x \right) \right> \left< \phi _1\left( y \right) \right> ,
\\
    \label{eq:46}
K_3\left( x,y \right) &=\left< \left( \phi _2\left( x \right) -\Phi _2\left( x \right) \right) \left( \chi \left( y \right) -\Phi _3\left( y \right) \right) \right> =\left< \phi _2\left( x \right) \chi \left( y \right) \right> -\left< \phi _2\left( x \right) \right> \left< \chi \left( y \right) \right> ,
\\
    \label{eq:47}
\bar{K}_3\left( x,y \right) &=\left< \left( \chi \left( x \right) -\Phi _3\left( x \right) \right) \left( \phi _2\left( y \right) -\Phi _2\left( y \right) \right) \right> =\left< \chi \left( x \right) \phi _2\left( y \right) \right> -\left< \chi \left( x \right) \right> \left< \phi _2\left( y \right) \right> .
\end{align}
At coincident spacetime points, we also define
\begin{equation}
    \label{eq:48}
G_i\left( x,x \right) =G_i\left( y,y \right) \equiv G_i, ~     K_i\left( x,x \right) =K_i\left( y,y \right) =\bar{K}_i\left( x,x \right) =\bar{K}_i\left( y,y \right) \equiv K_i=\bar{K}_i, ~ i=1,2,3 .
\end{equation}
These correlators contain ultraviolet divergences, which are regulated by the lattice discretization and absorbed into renormalized parameters below.

To close the equations of motion, we set all fully connected correlators of order three and higher to zero
\begin{equation}
    \label{eq:49}
\left< \phi _{i_1}\left( x_1 \right) \phi _{i_2}\left( x_2 \right) \cdots\phi _{i_n}\left( x_n \right) \right> _C=0, ~~ n\geqslant 3,
\end{equation}
where $i=1,2,3$, and the superscript $C$ denotes the fully connected part. With this Hartree truncation, all disconnected contributions can then be written in terms of $\Phi_i$ and $G_i, K_i, \bar{K_i}$. 

The dynamics are therefore truncated at the level of one- and two-point functions. In the quantum theory, Eqs.~\eqref{eq:35},~\eqref{eq:36} and~\eqref{eq:37} are promoted to Heisenberg equations of motion for the operators $\phi_1$, $\phi_2$, and $\chi$. They provide a starting point for the Schwinger-Dyson hierarchy for correlation functions. After the Hartree truncation, the hierarchy closes on the mean fields and connected two-point functions, which then determine the self-consistent evolution.

After the operator equations~\eqref{eq:35},~\eqref{eq:36} and~\eqref{eq:37} are averaged in the quantum state, the Hartree closure turns them into the following closed equations for the mean fields:
\begin{equation}
    \label{eq:50}
\left[ \partial _{x}^{2}+M_{1}^{2}\left( x \right) \right] \Phi _1\left( x \right) +M_{2}^{2}\left( x \right) \Phi _3\left( x \right) =0,
\end{equation}
\begin{equation}
    \label{eq:51}
\left[ \partial _{x}^{2}+M_{1}^{2}\left( x \right) \right] \Phi _2\left( x \right) +M_{3}^{2}\left( x \right) \Phi _3\left( x \right) =0,
\end{equation}
\begin{equation}
    \label{eq:52}
\left[ \partial _{x}^{2}+M_{4}^{2}\left( x \right) \right] \Phi _3\left( x \right) +M_{2}^{2}\left( x \right) \Phi _1\left( x \right) +M_{3}^{2}\left( x \right) \Phi _2\left( x \right) =0,
\end{equation}
where
\begin{equation}
    \label{eq:53}
M_{1}^{2} =h^2\left( \Phi _{3}^{2}+G_3 \right)+w_{1}^{2},
\end{equation}
\begin{equation}
    \label{eq:54}
M_{2}^{2} =2h^2K_2,
\end{equation}
\begin{equation}
    \label{eq:55}
M_{3}^{2} =2h^2K_3,
\end{equation}
\begin{equation}
    \label{eq:56}
M_{4}^{2} =h^2\left( \Phi _{1}^{2}+\Phi _{2}^{2}+G_1+G_2 \right) +\frac{1}{2}g^2\left( \Phi _{3}^{2}+3G_3-\chi _{v}^{2} \right)+w_{2}^{2}.
\end{equation}
If the two-point functions are omitted, these equations reduce to the form of the classical field equations. To obtain the equations for the two-point functions, we multiply the Heisenberg field equations by $\phi _i\left( y \right), \chi\left( y \right) $ and then take the quantum expectation values
\begin{equation}
    \label{eq:57}
\left[ \partial _{x}^{2}+\bar{M}_{1}^{2}\left( x \right) \right] G_1\left( x,y \right) +\bar{M}_{2}^{2}\left( x \right)\bar{K}_2\left( x,y \right) =0,
\end{equation}
\begin{equation}
    \label{eq:58}
\left[ \partial _{x}^{2}+\bar{M}_{1}^{2}\left( x \right) \right] K_1\left( x,y \right) +\bar{M}_{2}^{2}\left( x \right)\bar{K}_3\left( x,y \right) =0,
\end{equation}
\begin{equation}
    \label{eq:59}
\left[ \partial _{x}^{2}+\bar{M}_{1}^{2}\left( x \right) \right] K_2\left( x,y \right) +\bar{M}_{2}^{2}\left( x \right)G_3\left( x,y \right) =0,
\end{equation}
\begin{equation}
    \label{eq:60}
\left[ \partial _{x}^{2}+\bar{M}_{1}^{2}\left( x \right) \right] \bar{K}_1\left( x,y \right) +\bar{M}_{3}^{2}\left( x \right)\bar{K}_2\left( x,y \right) =0,
\end{equation}
\begin{equation}
    \label{eq:61}
\left[ \partial _{x}^{2}+\bar{M}_{1}^{2}\left( x \right) \right] G_2\left( x,y \right) +\bar{M}_{3}^{2}\left( x \right)\bar{K}_3\left( x,y \right) =0,
\end{equation}
\begin{equation}
    \label{eq:62}
\left[ \partial _{x}^{2}+\bar{M}_{1}^{2}\left( x \right) \right] K_3\left( x,y \right) +\bar{M}_{3}^{2}\left( x \right)G_3\left( x,y \right) =0,
\end{equation}
\begin{equation}
    \label{eq:63}
\left[ \partial _{x}^{2}+\bar{M}_{4}^{2}\left( x \right) \right] \bar{K}_2\left( x,y \right) +\bar{M}_{2}^{2}\left( x \right)G_1\left( x,y \right) +\bar{M}_{3}^{2}\left( x \right)\bar{K}_1\left( x,y \right) =0,
\end{equation}
\begin{equation}
    \label{eq:64}
\left[ \partial _{x}^{2}+\bar{M}_{4}^{2}\left( x \right) \right] \bar{K}_3\left( x,y \right) +\bar{M}_{2}^{2}\left( x \right)K_1\left( x,y \right) +\bar{M}_{3}^{2}\left( x \right)G_2\left( x,y \right) =0,
\end{equation}
\begin{equation}
    \label{eq:65}
\left[ \partial _{x}^{2}+\bar{M}_{4}^{2}\left( x \right) \right] G_3\left( x,y \right) +\bar{M}_{2}^{2}\left( x \right)K_2\left( x,y \right) +\bar{M}_{3}^{2}\left( x \right)K_3\left( x,y \right) =0,
\end{equation}
where
\begin{equation}
    \label{eq:66}
\bar{M}_{1}^{2}=h^2\left( \Phi _{3}^{2}+G_3 \right)+w_{1}^{2},
\end{equation}
\begin{equation}
    \label{eq:67}
\bar{M}_{2}^{2}=2h^2\left( K_2+\Phi _1\Phi _3 \right) ,
\end{equation}
\begin{equation}
    \label{eq:68}
\bar{M}_{3}^{2}=2h^2\left( K_3+\Phi _2\Phi _3 \right) ,
\end{equation}
\begin{equation}
    \label{eq:69}
\bar{M}_{4}^{2}=h^2\left( \Phi _{1}^{2}+\Phi _{2}^{2}+G_1+G_2 \right) +\frac{1}{2}g^2\left( 3\Phi _{3}^{2}+3G_3-\chi _{v}^{2} \right)+w_{2}^{2}.
\end{equation}

The two-point functions depend on both $x$ and $y$, so evolving them directly on the lattice is computationally expensive: each pair of lattice sites carries an independent unknown function. In a homogeneous background, either with vanishing mean fields or with spatially constant profiles, $\Phi _i\left( x \right) =\Phi _i\left( t \right) $ and $M, \bar{M}$ are also only time-dependent. The two-point correlation functions then simplify to
\begin{equation}
    \label{eq:70}
G_i\left( x,y \right) =G_i\left( t,t',|\mathbf{x}-\mathbf{y}| \right) , ~~ K_i\left( x,y \right) =K_i\left( t,t',|\mathbf{x}-\mathbf{y}| \right) , ~~ \bar{K}_i\left( x,y \right) =\bar{K}_i\left( t,t',|\mathbf{x}-\mathbf{y}| \right) ,
\end{equation}
which greatly reduces the number of degrees of freedom that must be evolved numerically.

Solitons, however, are spatially inhomogeneous. We take the soliton profiles as the initial mean fields and treat $\phi _i\left( x \right) -\Phi _i\left( x \right), \chi \left( x \right) -\Phi _3\left( x \right)  $ as Gaussian fluctuations. For convenience, we define 
\begin{equation}
\varphi _{1}\left( x \right) =\phi _{1}\left( x \right) -\Phi _{1}\left( x \right),\quad\varphi _{2}\left( x \right) =\phi _{2}\left( x \right) -\Phi _{2}\left( x \right),\quad \varphi _3\left( x \right) =\chi\left( x \right) -\Phi _3\left( x \right) .
\end{equation}
For Gaussian initial fluctuations, the perturbation fields can be expanded in time-dependent modes through the Bogoliubov transformation
\begin{equation}
    \label{eq:71}
\varphi _i\left( x \right) =\int{\widetilde{\mathrm{d}k_i}\left[ a_{\mathbf{k}}^{i}f_{\mathbf{k}}^{i}\left( x \right) +a_{\mathbf{k}}^{i\dagger}f_{\mathbf{k}}^{i*}\left( x \right) \right]}, \quad i=1,2,3,
\end{equation}
where $\widetilde{\mathrm{d}k_i}=\mathrm{d}^3k/\big[ \left( 2\pi \right) ^32\omega _{ki} \big]  $ with $\omega _{k1}=\omega _{k2}=({\mathbf{k}^2+m_{\phi r}^{2}})^{1/2}\,\,, \omega _{k3}=({\mathbf{k}^2+m_{\chi r}^{2}})^{1/2}$, and $m_{\phi r}^{2},m_{\chi r}^{2}$ are the renormalized masses to be defined later. The canonical commutation relations of the annihilation and creation operators $a_{\mathbf{k}}^{i}$ and $a_{\mathbf{k}}^{i\dagger}$ are
\begin{equation}
    \label{eq:72}
\left[ a_{\mathbf{k}}^{i}, a_{\mathbf{k}\prime}^{i'\dagger} \right] =\left( 2\pi \right) ^32\omega _{ki}\delta ^3\left( \mathbf{k}-\mathbf{k}' \right) \delta _{i,i'}, ~~ \left[ a_{\mathbf{k}}^{i}, a_{\mathbf{k}'}^{i'} \right] =\left[ a_{\mathbf{k}}^{i\dagger}, a_{\mathbf{k}'}^{i'\dagger} \right] =0.
\end{equation}
It follows that the mode functions $f_{\mathbf{k}}^{i}$ obey equations with the same structure as the equations for the two-point functions
\begin{equation}
\label{eq:mode1}
    \left[ \partial _{x}^{2}+\bar{M}_{1}^{2}\left( x \right) \right] f_{\mathbf{k}}^{1}\left( x \right) +\bar{M}_{2}^{2}\left( x \right) f_{\mathbf{k}}^{3}\left( x \right) =0,
\end{equation}
\begin{equation}
\label{eq:mode2}
\left[ \partial _{x}^{2}+\bar{M}_{1}^{2}\left( x \right) \right] f_{\mathbf{k}}^{2}\left( x \right) +\bar{M}_{3}^{2}\left( x \right) f_{\mathbf{k}}^{3}\left( x \right) =0,
\end{equation}
\begin{equation}
\label{eq:mode3}
\left[ \partial _{x}^{2}+\bar{M}_{4}^{2}\left( x \right) \right] f_{\mathbf{k}}^{3}\left( x \right) +\bar{M}_2^2\left( x \right) f_{\mathbf{k}}^{1}\left( x \right) +\bar{M}_{3}^{2}\left( x \right) f_{\mathbf{k}}^{2}\left( x \right) =0.
\end{equation}

For a quantum state containing the vacuum state associated with the initial mean-field background, the two-point functions are given by
\begin{equation}
G_i=\int \widetilde{\mathrm{d}k_i}\,
\left|f_{\mathbf{k}}^i(x)\right|^2,
\qquad
K_i=0 .
\end{equation}
In principle, the quantum system can therefore be evolved by integrating the mode functions, computing $G_i$, and evolving the mean fields $\Phi _i$ self-consistently. This procedure is expensive because one mode function is required for each Fourier mode, and the number of modes scales as the huge number of lattice sites on the $3$D lattice. We therefore replace the full mode-function evolution by a stochastic ensemble representation, described next.

\subsection{Stochastic Ensemble Average}
\label{sec:ensemble}
The stochastic ensemble method \cite{Borsanyi:2007wm} replaces the explicit quantum mode functions by stochastic classical fields that reproduce the same two-point functions. We define the stochastic field $\varphi _{i}^{e}\left( x \right) $ as
\begin{equation}
    \label{eq:74}
\varphi _{i}^{e}\left( x \right) =\int{\widetilde{dk}_i\left[ c_{\mathbf{k}}^{i,e}f_{\mathbf{k}}^{i}\left( x \right) +c_{\mathbf{k}}^{i,e*}f_{\mathbf{k}}^{i*}\left( x \right) \right]}, \quad i=1,2,3,
\end{equation}
where $e$ labels the ensemble realization and $c_{\mathbf{k}}^{i,e}$ is a Gaussian random coefficient with vanishing mean and variance
\begin{equation}
    \label{eq:75}
\left< c_{\mathbf{k}}^{i,e*}c_{\mathbf{k}'}^{i,e} \right> _E=\left( 2\pi \right) ^3\omega _{ki}\delta ^3\left( \mathbf{k}-\mathbf{k}' \right) .
\end{equation}
Here $E$ denotes the ensemble average over $e$. The equations of motion for $\varphi _{i}^{e}\left( x \right) $ have the same form as the mode-function equations
\begin{equation}
\label{eq:sto1}
\left[ \partial _{x}^{2}+\bar{M}_{1}^{2}\left( x \right) \right] \varphi _{1}^{e}\left( x \right) +\bar{M}_{2}^{2}\left( x \right) \varphi _{3}^{e}\left( x \right) =0,
\end{equation}
\begin{equation}
\label{eq:sto2}
\left[ \partial _{x}^{2}+\bar{M}_{1}^{2}\left( x \right) \right] \varphi _{2}^{e}\left( x \right) +\bar{M}_{3}^{2}\left( x \right) \varphi _{3}^{e}\left( x \right) =0,
\end{equation}
\begin{equation}
\label{eq:sto3}
\left[ \partial _{x}^{2}+\bar{M}_{4}^{2}\left( x \right) \right] \varphi _{3}^{e}\left( x \right) +\bar{M}_2^2\left( x \right) \varphi _{1}^{e}\left( x \right) +\bar{M}_{3}^{2}\left( x \right) \varphi _{2}^{e}\left( x \right) =0.
\end{equation}

By construction, the stochastic average reproduces the quantum two-point correlators in the infinite-ensemble limit
\begin{equation}
    \label{eq:76}
\left< \varphi _{i}^{e}\left( x \right) ^2 \right> _E-\left< \varphi _{i}^{e}\left( x \right) \right> _{E}^{2}=G_i,
\end{equation}
\begin{equation}
    \label{eq:77}
\left< \varphi _{1}^{e}\left( x \right) \varphi _{2}^{e}\left( x \right) \right> _E-\left< \varphi _{1}^{e}\left( x \right) \right> _E\left< \varphi _{2}^{e}\left( x \right) \right> _E=K_1=\bar{K}_1,
\end{equation}
\begin{equation}
    \label{eq:78}
\left< \varphi _{1}^{e}\left( x \right) \varphi _{3}^{e}\left( x \right) \right> _E-\left< \varphi _{1}^{e}\left( x \right) \right> _E\left< \varphi _{3}^{e}\left( x \right) \right> _E=K_2=\bar{K}_2,
\end{equation}
\begin{equation}
    \label{eq:79}
\left< \varphi _{2}^{e}\left( x \right) \varphi _{3}^{e}\left( x \right) \right> _E-\left< \varphi _{2}^{e}\left( x \right) \right> _E\left< \varphi _{3}^{e}\left( x \right) \right> _E=K_3=\bar{K}_3.
\end{equation}
Thus the quantum expectation values can be replaced by ensemble averages over stochastic fields. In practice, we evolve the one-point equations together with many independent stochastic realizations and then average over the ensemble to obtain the two-point functions. When the number of realizations needed for good statistics is much smaller than the number of mode functions, or equivalently the number of lattice sites in $3$ spatial dimensions, $\mathcal{E} \ll N^3$, this procedure gives a substantial computational speed-up. 

\subsection{Renormalization and the Classical Limit}
\label{sec:Renormalization}
The two-point functions are formally divergent at coincident points. On the lattice, these divergences become finite but cutoff-dependent quantities, which must be absorbed into the parameters of the Lagrangian through renormalization. 

We choose the initial mode functions as
\begin{equation}
    \label{eq:80}
f_{\mathbf{k}}^{i}\left( x \right) =e^{\mathrm{i}kx}, \qquad\partial _tf_{\mathbf{k}}^{i}\left( x \right) =-\mathrm{i}\omega _{ki}e^{\mathrm{i}kx}.
\end{equation}
The initial two-point functions are then
\begin{align}
    \label{eq:81}
G_{1}^{\left( \text{initial} \right)}\left( x,x \right) =\int{\frac{\mathrm{d}^3k}{\left( 2\pi \right) ^32\omega _{k1}}}\equiv A, 
\\
G_{2}^{\left(  \text{initial} \right)}\left( x,x \right) =\int{\frac{\mathrm{d}^3k}{\left( 2\pi \right) ^32\omega _{k2}}}\equiv A, 
\\
G_{3}^{\left(  \text{initial} \right)}\left( x,x \right) =\int{\frac{\mathrm{d}^3k}{\left( 2\pi \right) ^32\omega _{k3}}}\equiv B.
\end{align}
These quantities exhibit quadratic and logarithmic divergences in $3+1$ dimensions. We implement the renormalization by introducing the renormalized parameters $h_r, g_r, \chi_{vr}, w_{1r}, w_{2r}$
\begin{align}
    \label{eq:82}
h_r&=h,
\\
    \label{eq:83}
g_r&=g,
\\
    \label{eq:84}
w_{1r}^{2}&=w_{1}^{2}+h^2B=0,
\\
    w_{2r}^2&=w_2^2+2h^2A=0,
\\
    \chi _{vr}^{2}&=\chi _{v}^{2}-3B.
\end{align}
We identify the renormalized parameters with the parameters in the classical action and set $w_{1r}=w_{2r}=0$. With these conditions, the equations of motion smoothly match the classical dynamics, with the classical parameters replaced by their renormalized values, at the initial time. The renormalized field masses in the definitions of $\omega_{ki}$ are therefore $m_{\chi r}=g_r\chi_{vr},m_{\phi r}=h_r\chi_{vr}$. 

These renormalization conditions also allow us to study the classical limit of the FLS model. This limit corresponds to small couplings, or equivalently to large-amplitude Q-balls \cite{Coleman:1985rnk} \footnote{Under this scaling, $\chi _v$ also changes, so the frequency $\omega$ changes accordingly. However, $\omega/h$ remains fixed throughout the scaling. In the quantum simulations below, when we say that the dimensionless ratio $\omega/h$ is kept fixed while varying the scaling parameter, we mean that $\omega/h$ is kept fixed.} 
\begin{equation}
    \label{eq:90}
\Phi_i\rightarrow \Phi_i/\sqrt{\alpha}, \quad g^2\rightarrow \alpha g^2, \quad {\chi}^2_{v}\rightarrow {\chi}^2_{v}/\alpha , \quad h^2\rightarrow \alpha h^2.
\end{equation}
The scaling factor $\alpha$ is a dimensionless real constant, and the classical limit is $\alpha\to0$. The size of the quantum corrections is therefore controlled by this scaling parameter. This can be seen explicitly by writing the mean-field equations in terms of renormalized parameters
\begin{align}
( \partial _{x}^{2}+w_{1r}^{2}&+h_{r}^{2}\Phi_3 ^2 ) \Phi_1 =h_{r}^{2}\left( B-G_3 \right) \Phi_1-2h_r^2\Phi_3K_2,
\\
( \partial _{x}^{2}+w_{1r}^{2}&+h_{r}^{2}\Phi_3 ^2 ) \Phi_2 =h_{r}^{2}\left( B-G_3 \right) \Phi_2-2h_r^2\Phi_3K_3,
\\
( \partial _{x}^{2}+w_{2r}^{2}&+h_{r}^{2}(\Phi_1 ^2+\Phi_2^2)+\frac{1}{2}g_{r}^{2}\Phi_3 ^2-\frac{1}{2}g_{r}^{2}\chi _{vr}^{2} ) \Phi_3 = 
\\
&h_{r}^{2}\left( A-G_1+A-G_2 \right) \Phi_3 +\frac{3}{2}g_{r}^{2}\left( B-G_3 \right) \Phi_3 -2h_r^2(\Phi_1K_2+\Phi_2K_3). \notag
\end{align}
Substituting Eq.~\eqref{eq:90} into these equations, we obtain
\begin{align}
( \partial _{x}^{2}+w_{1r}^{2}+ & h_{r}^{2}\Phi_3 ^2 ) \Phi_1 =\alpha\left[h_{r}^{2}\left( B-G_3 \right) \Phi_1-2h_r^2\Phi_3K_2\right] ,
\\
( \partial _{x}^{2}+w_{1r}^{2}+ & h_{r}^{2}\Phi_3 ^2 ) \Phi_2 =\alpha\left[h_{r}^{2}\left( B-G_3 \right) \Phi_2-2h_r^2\Phi_3K_3\right] ,
\\
( \partial _{x}^{2}+w_{2r}^{2}+ & h_{r}^{2}(\Phi_1 ^2+\Phi_2^2)+\frac{1}{2}g_{r}^{2}\Phi_3 ^2-\frac{1}{2}g_{r}^{2}\chi _{vr}^{2} ) \Phi_3 =
\\
&\alpha\left[h_{r}^{2}\left( A-G_1+A-G_2 \right) \Phi_3 +\frac{3}{2}g_{r}^{2}\left( B-G_3 \right) \Phi_3 -2h_r^2(\Phi_1K_2+\Phi_2K_3)\right].\nonumber
\end{align}
Under this scaling the classical part of the equations is unchanged, whereas every Hartree
source term acquires an overall factor of \(\alpha\). Hence the rescaled theory contains a
solution with the same dimensionless profile, but with field amplitudes larger by
\(1/\sqrt{\alpha}\). Equivalently, one may keep the original profile fixed and regard the
Hartree backreaction as being suppressed according to
\[
A-G_1\mapsto \alpha(A-G_1),\qquad
A-G_2\mapsto \alpha(A-G_2),\qquad
B-G_3\mapsto \alpha(B-G_3).
\]
The parameter \(\alpha\) therefore provides a direct handle on the relative importance of
quantum fluctuations: smaller \(\alpha\) corresponds to a more classical, larger-amplitude
configuration.

\subsection{Numerical Quantities and Observables}
\label{sec:Numerical}
To compute the real-time dynamics on the lattice, we define dimensionless quantities
\begin{equation}
    \label{eq:87}
\tilde{x}^{\mu}=\chi _{vr}x^{\mu}, ~  \tilde{\chi}=\chi /\chi _{vr}, ~  \tilde{\phi}=\phi /\chi _{vr}, ~ \tilde{g}=g_{r}, ~ \tilde{h}=h_{r},  \tilde{w}_{1}=w_{1r}/\chi _{vr}, ~ \tilde{w}_{2}=w_{2r}/\chi _{vr},
\end{equation}
such that all quantities are cast in units of $\chi_{vr}$, the renormalized vacuum expectation value of the real field. Using these variables, the classical action, for example, becomes
\begin{equation}
    \label{eq:88}
S=\int{\mathrm{d}^4\tilde{x}\left[ |\tilde{\partial}_{\mu}\tilde{\phi}|^2+\frac{1}{2}\left( \tilde{\partial}_{\mu}\tilde{\chi} \right) ^2-\tilde{h}^2\tilde{\chi}^2|\tilde{\phi}|^2-\frac{\tilde{g}^2}{8}\left( \tilde{\chi}^2-\,\,1 \right) ^2-\tilde{w}_{1}^{2}|\tilde{\phi}|^2-\frac{1}{2}\tilde{w}_{2}^{2}\tilde{\chi}^2 \right]},
\end{equation}
which now depends on four free parameters. 

We focus on $3+1$D simulations. The lattice spacing is chosen as $\chi_{vr}\mathrm{d}x=0.5$, so the highest lattice momentum is $k_{\max}/\chi_{vr}=\sqrt{3}\pi /\left( \chi_{vr}\mathrm{d}x \right) \sim O\left( 10 \right) $. The ultraviolet cutoff is therefore about one order of magnitude above the scale $\chi_{vr}$. The code is based on the LATfield2 package \cite{Daverio:2015ryl}, which allows straightforward parallel computing of field evolution. Spatial gradients are discretized with a second-order finite-difference operator, and the fields are advanced in time using a second-order leapfrog integrator. Spatial boundaries are periodic; they do not affect our results because the stable Q-balls studied here are approximately stationary and radiate only weakly.

For the mean fields, we use the classical soliton profiles as initial conditions. This choice ensures that the quantum evolution initially matches the classical evolution smoothly. However, the classical profiles are close to, but not exactly, soliton solutions of the quantum-corrected equations. This mismatch provides a small perturbation, which we use below to probe stability of quantum solitons. 

We quantify the difference between quantum-corrected and purely classical dynamics by considering the Noether charge and other observables. The total charge splits naturally into a mean-field contribution plus a fluctuation contribution

\begin{equation}
    \label{eq:92}
Q=\left< \hat{Q} \right> =Q_{\Phi}\left( t \right) +Q_G\left( t \right) ,
\end{equation}
where
\begin{equation}
    \label{eq:93}
Q_{\Phi}\left( t \right) =\int{\mathrm{d}^3x}\left( \Phi _1\partial _t\Phi _2-\Phi _2\partial _t\Phi _1 \right)  ,
\end{equation}
\begin{equation}
    \label{eq:94}
Q_G(t)=\int \mathrm{d}^3 x\,
\left<
\varphi_1\partial_t\varphi_2-\varphi_2\partial_t\varphi_1
\right>.
\end{equation}
We emphasize that this decomposition is a diagnostic split rather than a
decomposition into two separately conserved Noether charges.  The exact
$U(1)$ symmetry constrains only the total charge $Q$, while the mean-field
and fluctuation contributions may transfer charge to each other during the evolution.

The charge density likewise decomposes into a mean-field part and a fluctuation part
\begin{equation}
    \label{eq:95}
j^0=j_{\Phi}^{0}+j_{G}^{0}.
\end{equation}
In the simulations, the quantum average in Eq.~\eqref{eq:94} is replaced by an ensemble average over the stochastic fields $\varphi _{i}^{e}$. 

\section{Numerical Results}
In this section, we present our numerical simulations of Q-ball dynamics in the FLS model. We begin with the classical evolution in order to establish the basic properties of the soliton and to provide a reference for identifying genuine quantum effects. We then turn to the quantum dynamics within the inhomogeneous Hartree approximation, and study both the regime where the evolution remains close to the classical one and the regime where quantum fluctuations play an essential role. Finally, we analyze the stability of quantum-corrected Q-balls and examine how it depends on the model parameters. All quantities shown in this section are dimensionless and are therefore denoted with a tilde\footnote{For convenience, we omit the tildes on the frequency $\omega$ and the coupling constants $g,h$ , which are the main tuning parameters in the simulation, but the reader should note that they are also dimensionless in this section.}.

\label{sec:Results}
\subsection{Classical Dynamics}

As a classical baseline for the later quantum simulations, we first evolve an isolated Q-ball without Hartree fluctuations. We focus on the $3+1$D case and use $N=256$ lattice sites in each spatial direction, with spacings $\mathrm{d}\tilde{x}=\mathrm{d}\tilde{y}=\mathrm{d}\tilde{z}=0.5$ and $\mathrm{d}\tilde{t}=0.01$, which are sufficient for convergence. The parameters are $\omega=0.5h$ and $g=h=0.1$ (equivalently, $\alpha=0.01$ if $g=h=1$). Other parameter choices give qualitatively similar behavior, so this representative example captures the main features of the classical dynamics. 

\begin{figure}[tbp]
\centering
\includegraphics[width=.44\textwidth]{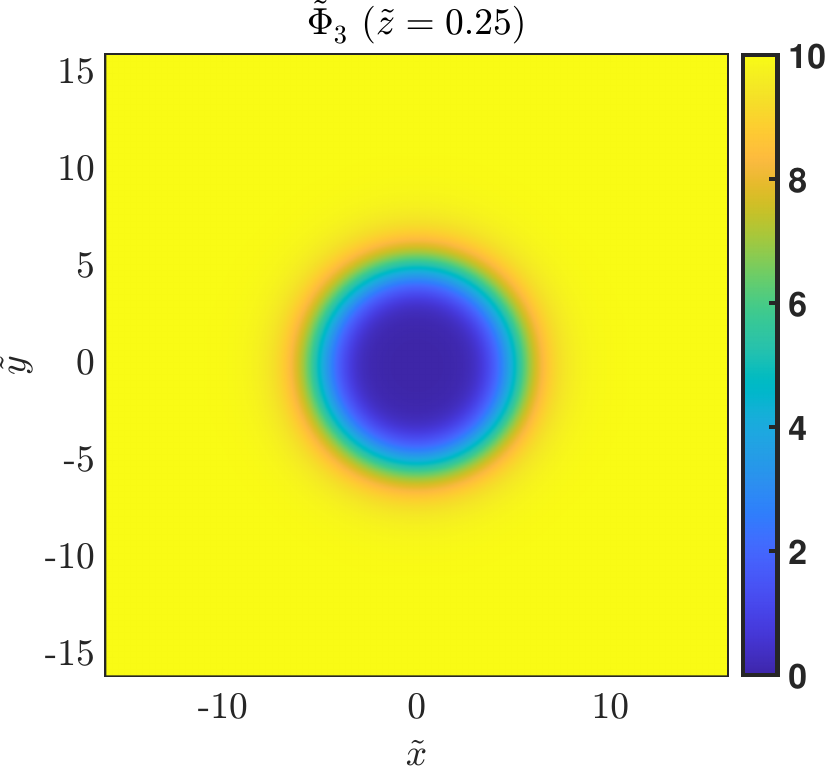}
\hfill
\includegraphics[width=.44\textwidth]{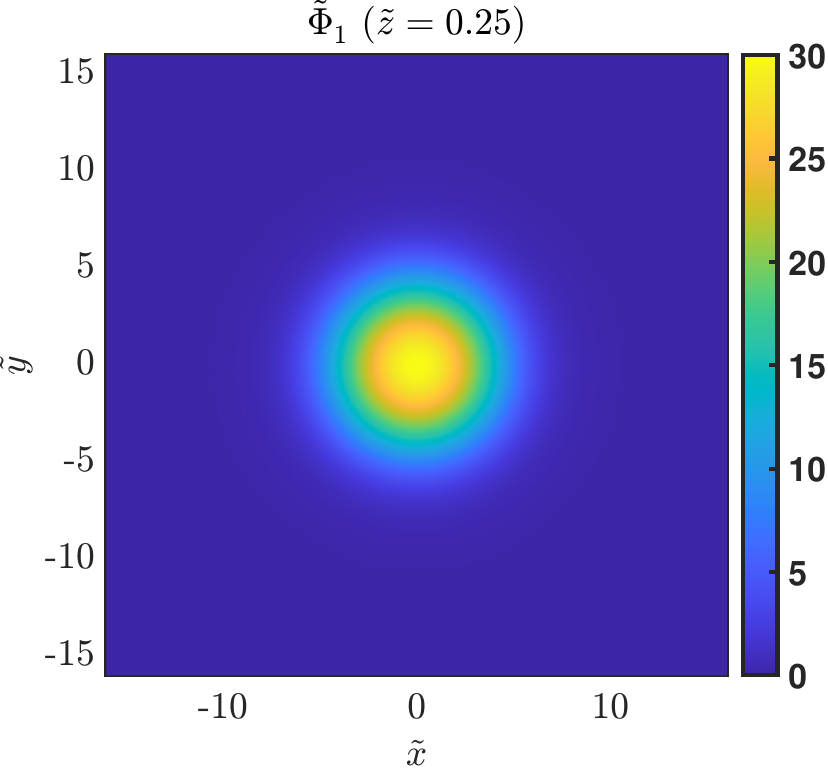}
\hfill
\includegraphics[width=.44\textwidth]{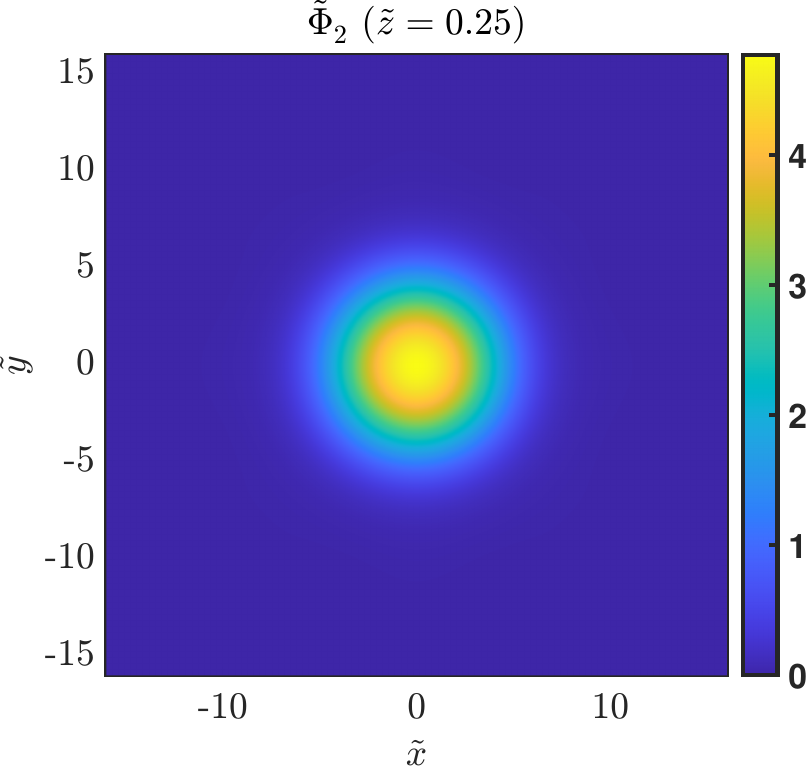}
\hfill
\includegraphics[width=.46\textwidth]{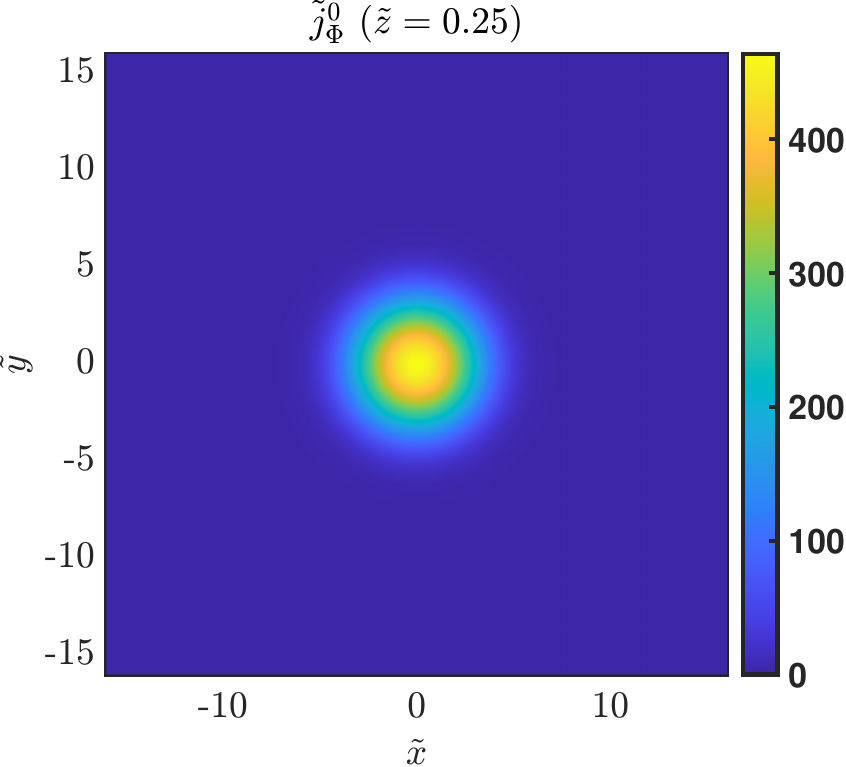}
\caption{Classical fields and charge density in a two-dimensional cross section at $\tilde{z}=0.25$ and $\tilde{t}=50$ in the classical simulation ($\omega=0.5h,~g=h=0.1$, with $\alpha=0.01$).\label{fig:6}}
\end{figure}

The classical FLS soliton is spatially localized and time-periodic, as in Eq.~\eqref{eq:6}. Since the configuration is spherically symmetric, two-dimensional density plots are sufficient to illustrate its structure without loss of essential information. Figure~\ref{fig:6} shows two-dimensional cross sections of the classical fields, which will later be compared with the quantum mean fields, together with the charge density. Outside the soliton, $\tilde{\Phi}_3$ approaches one of the degenerate vacua. Inside the soliton, with radius $R\sim5$, it deviates significantly from the vacuum. The fields $\tilde{\Phi}_1$ and $\tilde{\Phi}_2$ behave similarly: their vacuum value is zero, but their amplitudes become nonzero inside the soliton. The resulting charge density is localized and time-independent. 

\begin{figure}[tbp]
\centering
\includegraphics[width=.5\textwidth]{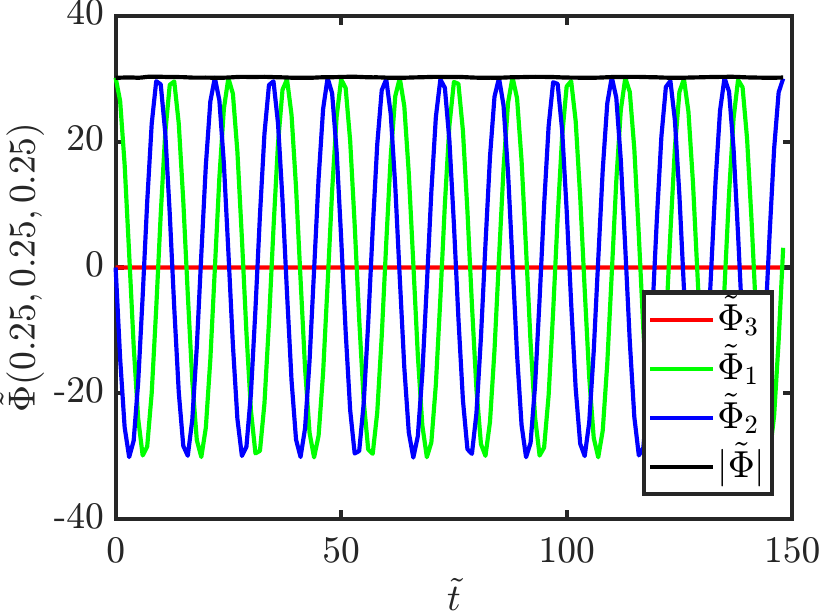}
\\
\includegraphics[width=.44\textwidth]{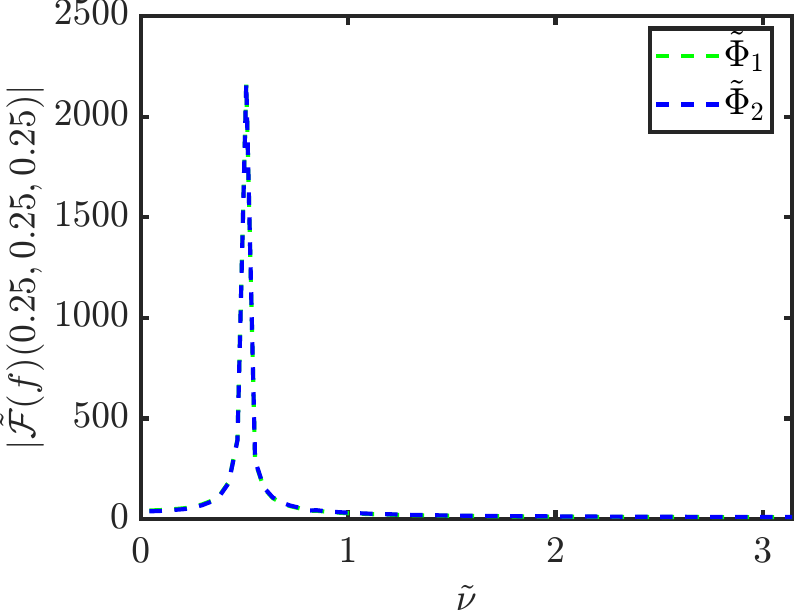}
\hfill
\includegraphics[width=.35\textwidth]{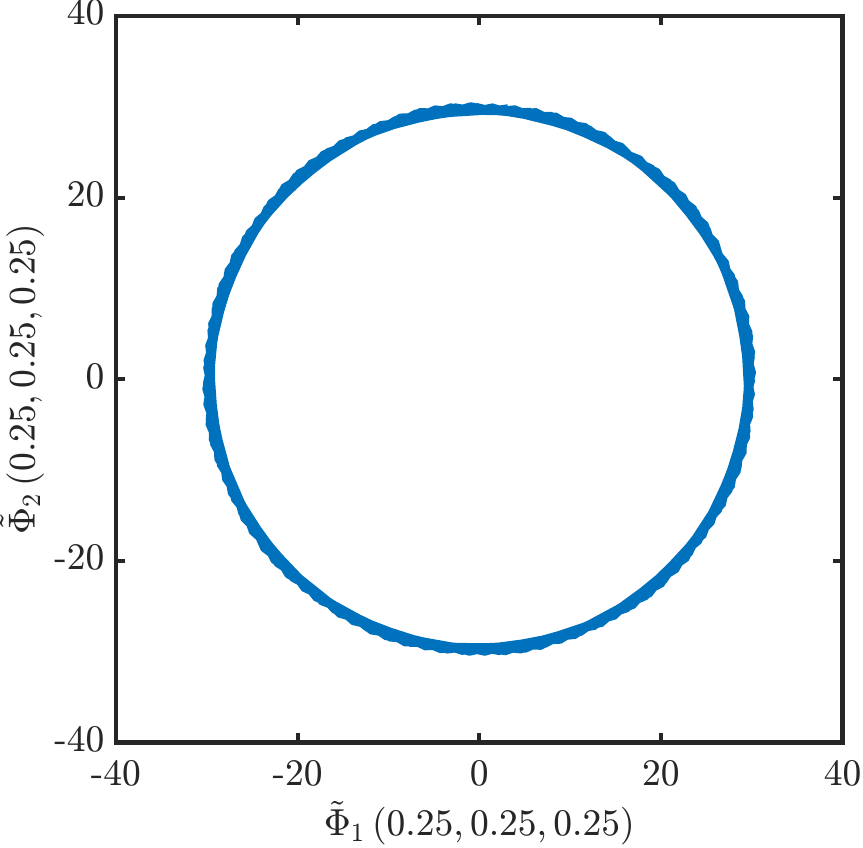}
\caption{Evolution of one-point functions (where $|\tilde{\Phi}|=\sqrt{\tilde{\Phi}_{1}^{2}+\tilde{\Phi}_{2}^{2}}$ is the modulus), their Fourier spectra and trajectory in the $\tilde{\Phi}_1-\tilde{\Phi}_2$ plane at the point $\left( \tilde{x},\tilde{y},\tilde{z} \right) =\left( 0.25,0.25,0.25 \right)$ in the classical simulation ($\omega=0.5h,~g=h=0.1$, with $\alpha=0.01$). The slight broadening of the Fourier peak is a numerical artifact due to the finite simulation time and lattice discretization; it serves as a reference for comparison with the Fourier spectra in the classical regime of the quantum simulation shown in figure \ref{fig:10}.\label{fig:7}}
\end{figure}

The first panel of figure~\ref{fig:7} shows the time evolution of the fields at the point $\left( \tilde{x},\tilde{y},\tilde{z} \right) =\left( 0.25,0.25,0.25 \right)$, near the soliton center. Both $\tilde{\Phi}_1$ and $\tilde{\Phi}_2$ oscillate sinusoidally with the same frequency, and their modulus remains constant in time, while $\tilde{\Phi}_3$ is static. The second panel shows the Fourier transforms of $\tilde{\Phi}_1$ and $\tilde{\Phi}_2$, both of which peak at $0.5m_{\phi r}$, exactly the input frequency. The slight broadening of the Fourier peak is a numerical artifact due to
the finite simulation time and lattice discretization; it serves as a reference for comparison with the
Fourier spectra in the classical regime of the quantum simulation shown later. The third panel shows the trajectory in the $\tilde{\Phi}_1-\tilde{\Phi}_2$ plane, which repeatedly traces a nearly perfect circle. 

\subsection{Quantum Dynamics}
We now turn to the quantum dynamics of Q-balls. We again work in $3+1$D and use $\mathrm{d}\tilde{x}=\mathrm{d}\tilde{y}=\mathrm{d}\tilde{z}=0.5$ and $\mathrm{d}\tilde{t}=0.01$, as in the classical simulations. The number of lattice points per spatial direction is $N=64$, and the number of stochastic realizations is $\mathcal{E} =15000$, which is sufficient for convergence. We classify the observed behavior into a classical regime (section~\ref{sec:classicalRegime}) and a quantum regime (section~\ref{sec:quantumRegime}), and then study the stability of quantum-corrected Q-balls (section~\ref{sec:stability}). 

\subsubsection{Classical Regime}
\label{sec:classicalRegime}

\begin{figure}[tbp]
\centering
\includegraphics[width=.46\textwidth]{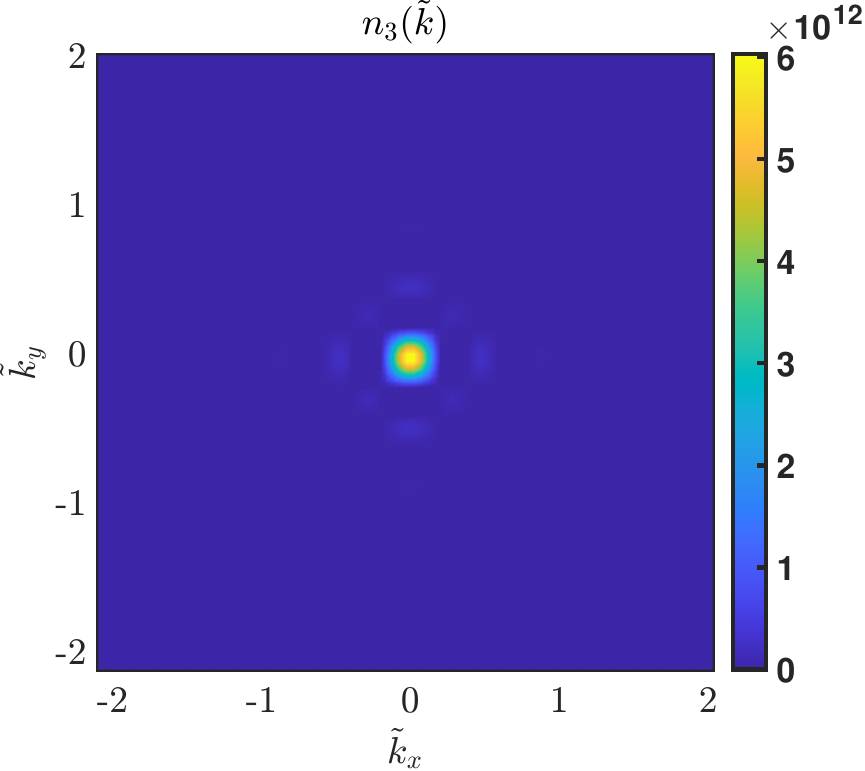}
\hfill
\includegraphics[width=.45\textwidth]{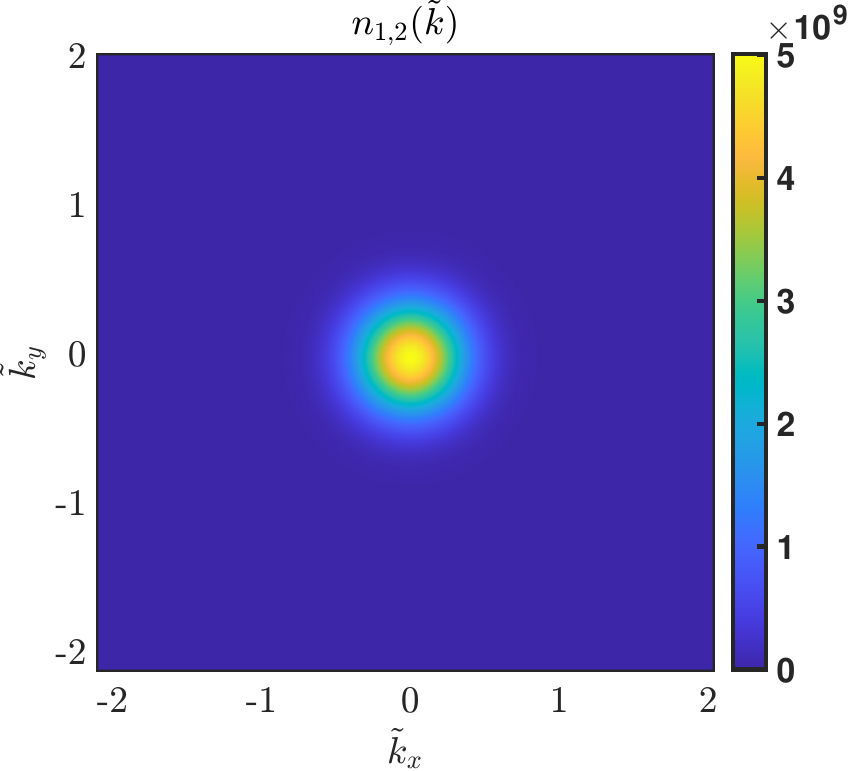}
\caption{Quantum occupation number distributions of Q-balls with $\omega=0.5h$ in a two-dimensional cross section at $\tilde{k}_z=\pi/32$ in the classical regime of the quantum dynamics ($g=h=0.1$, with $\alpha=0.01$). \label{fig:9}}
\end{figure}

The classical approximation is expected to be reliable when all relevant momentum modes are highly occupied. To investigate this, we use the Fourier power spectrum as a proxy for the occupation of momentum modes
\begin{equation}
  n_{1,2}(\tilde{t},\tilde{k})=\left| \int{\mathrm{d}^3\tilde{\mathbf{x}}\frac{1}{\sqrt{2}}\left( \tilde{\Phi}_1\left( \tilde{t},\tilde{\mathbf{x}} \right) +\mathrm{i}\tilde{\Phi}_2\left( \tilde{t},\tilde{\mathbf{x}} \right) \right) e^{-\mathrm{i}\tilde{k}\cdot \tilde{\mathbf{x}}}} \right|^2,    \quad n_3(\tilde{t},\tilde{k})=\left| \int{\mathrm{d}^3\tilde{\mathbf{x}}\tilde{\Phi}_3\left( \tilde{t},\tilde{\mathbf{x}} \right) e^{-\mathrm{i}\tilde{k}\cdot \tilde{\mathbf{x}}}} \right|^2.
\end{equation}
For Q-balls, these quantities are time-independent. Figure~\ref{fig:4} shows that low-frequency, thin-wall Q-balls carry large charges. The Fourier spectra in figure~\ref{fig:9} show that these large Q-balls also have much larger occupation numbers than small Q-balls; for example, $n_{3} (\tilde{k} )\sim 10^{11}$ and $n_{1,2} (\tilde{k} )\sim 10^{7}$ when $\omega=0.9h, ~g=h=0.2$, with $\alpha=0.04$. We now verify that, for such large Q-balls, the classical dynamics is indeed a good approximation.

\begin{figure}[tbp]
\centering
\includegraphics[width=.44\textwidth]{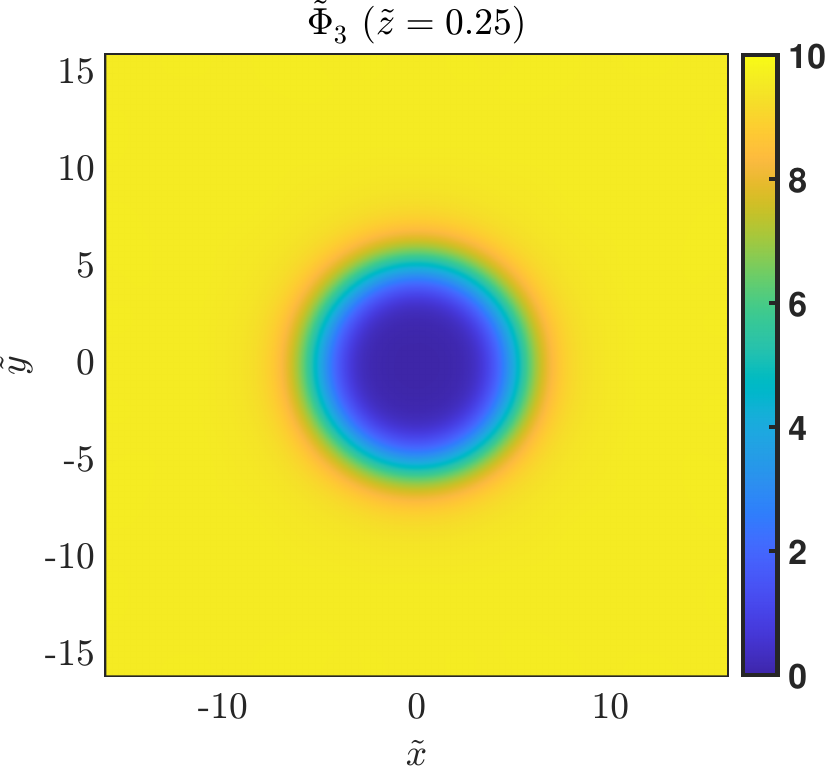}
\hfill
\includegraphics[width=.44\textwidth]{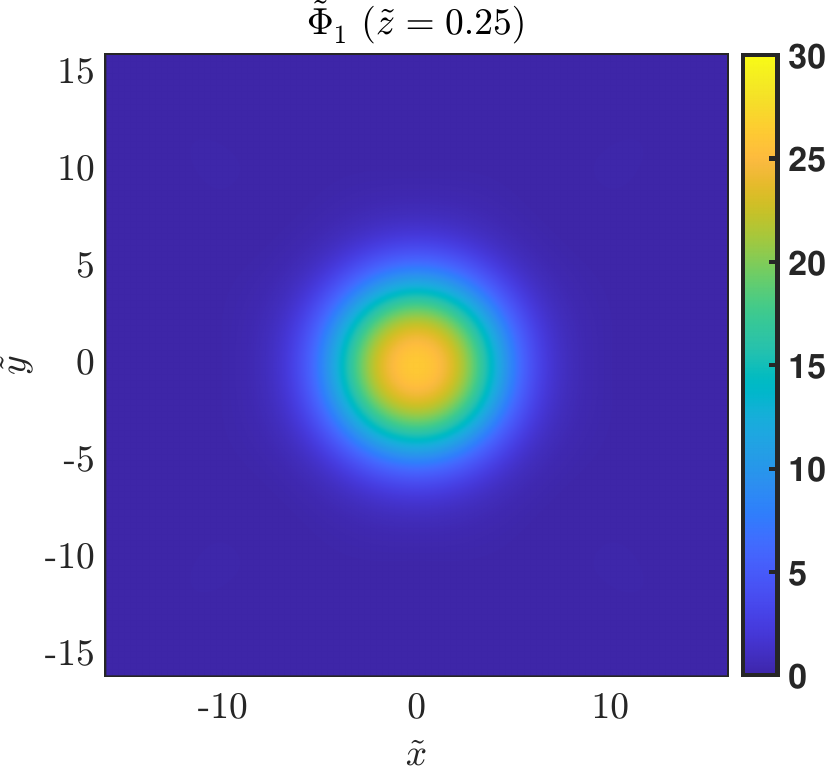}
\hfill
\includegraphics[width=.46\textwidth]{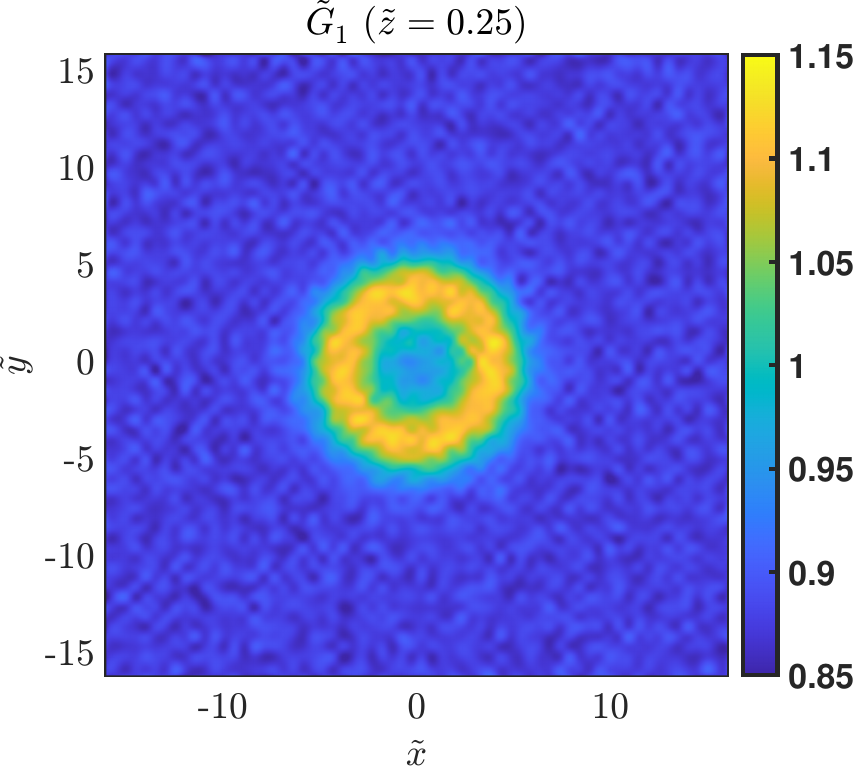}
\hfill
\includegraphics[width=.45\textwidth]{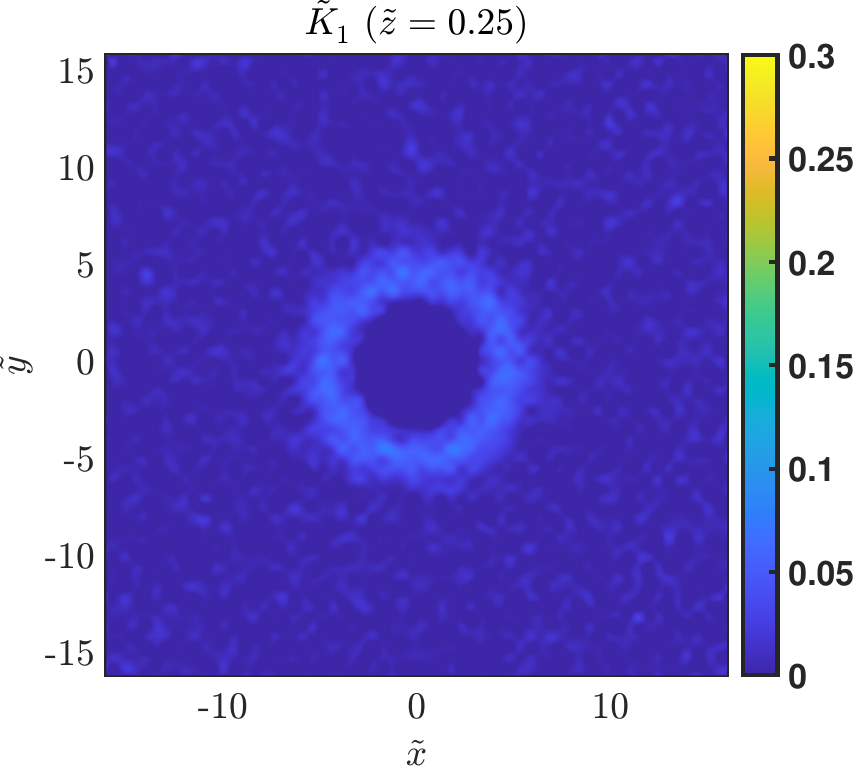}
\hfill
\includegraphics[width=.46\textwidth]{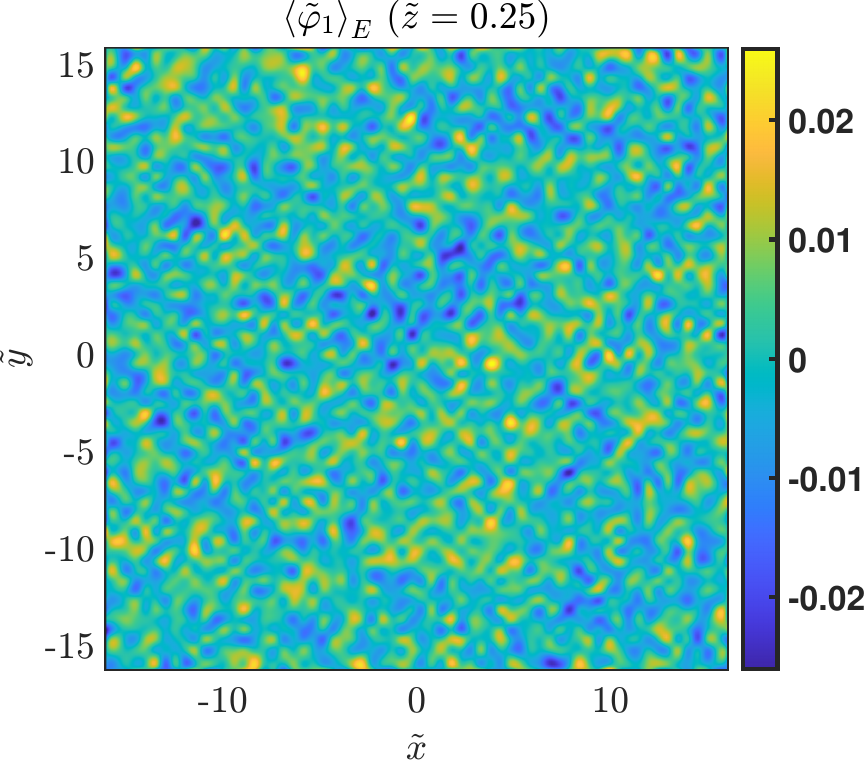}
\hfill
\includegraphics[width=.45\textwidth]{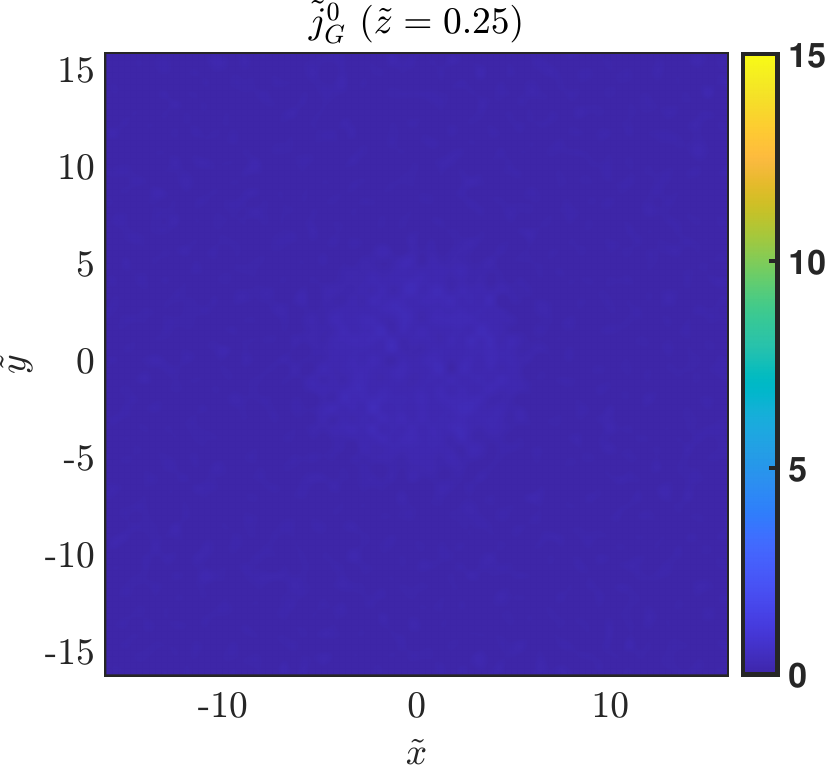}
\caption{Quantum-corrected density of $\tilde{\Phi}_i$, $\tilde{G}_i$, $\tilde{K}_i$, $\left< \tilde{\varphi} _i \right> _E$ and $\tilde{j}_{G}^{0}$ in a two-dimensional cross section at $\tilde{z}=0.25$ and $\tilde{t}=50$ in the classical regime of the quantum simulation ($\omega=0.5h,~g=h=0.1$, with $\alpha=0.01$).\label{fig:8}}
\end{figure}

We set $\omega=0.5h,~g=h=0.1$ (with scaling parameter $\alpha=0.01$ if $g=h=1$). Figure~\ref{fig:8} shows the densities of several representative quantities. The one-point functions $\tilde{\Phi}_i$ remain close to their classical counterparts (cf. figure~\ref{fig:6}). By contrast, the two-point functions $\tilde{G}_1$ and $\tilde{K}_1$ have small amplitudes and are homogeneous outside the Q-ball. The quantum contribution to the charge, $\tilde{j}_G^0$, is essentially unexcited and negligible compared with the mean-field contribution. The ensemble average $\left< \tilde{\varphi _i}\right> _E$ vanishes everywhere up to small statistical fluctuations, providing a useful check of the stochastic simulation. 

\begin{figure}[tbp]
\centering
\includegraphics[width=.44\textwidth]{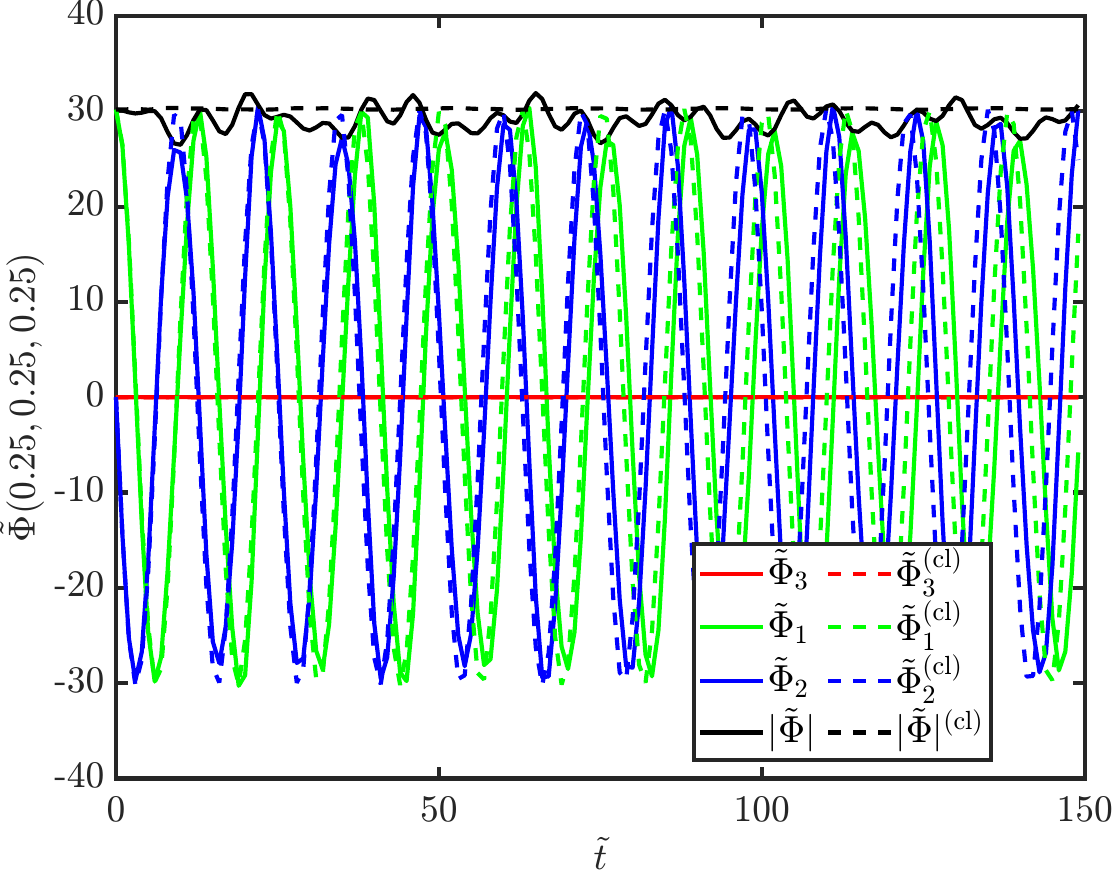}
\hfill
\includegraphics[width=.44\textwidth]{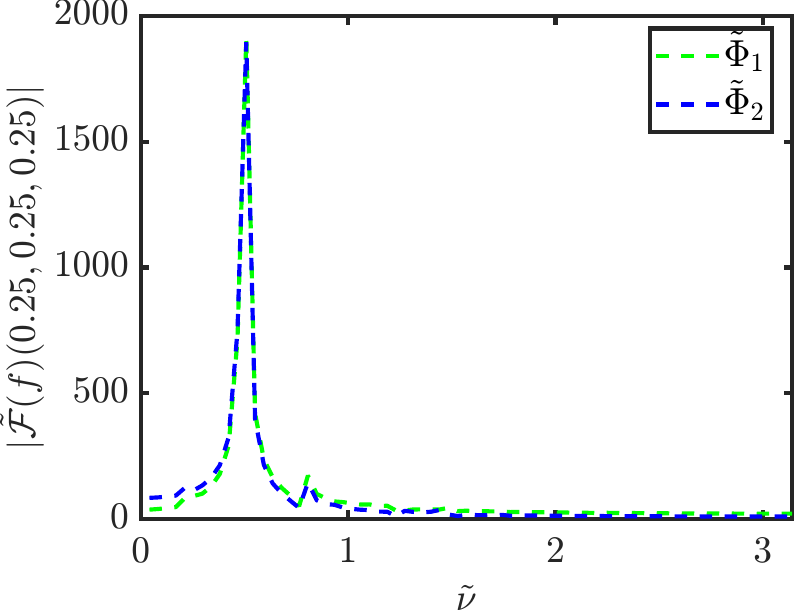}
\hfill
\includegraphics[width=.44\textwidth]{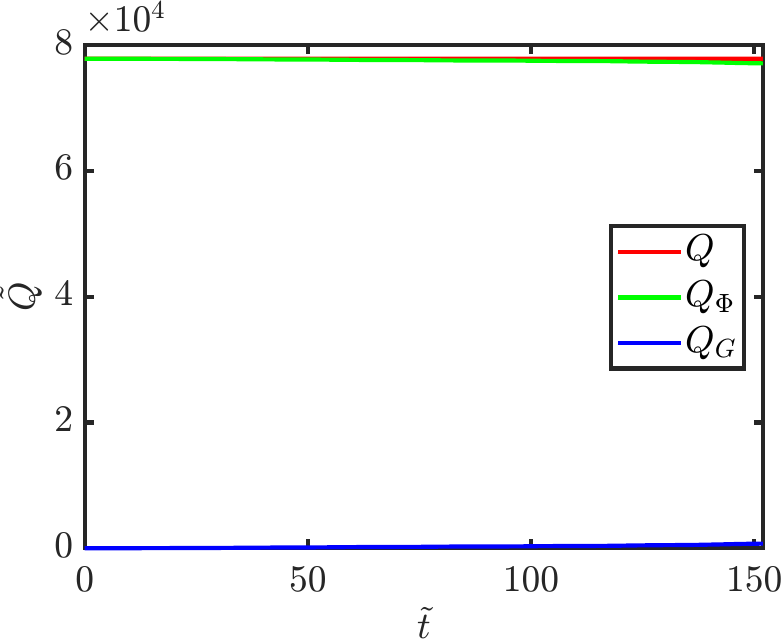}
\hfill
\includegraphics[width=.44\textwidth]{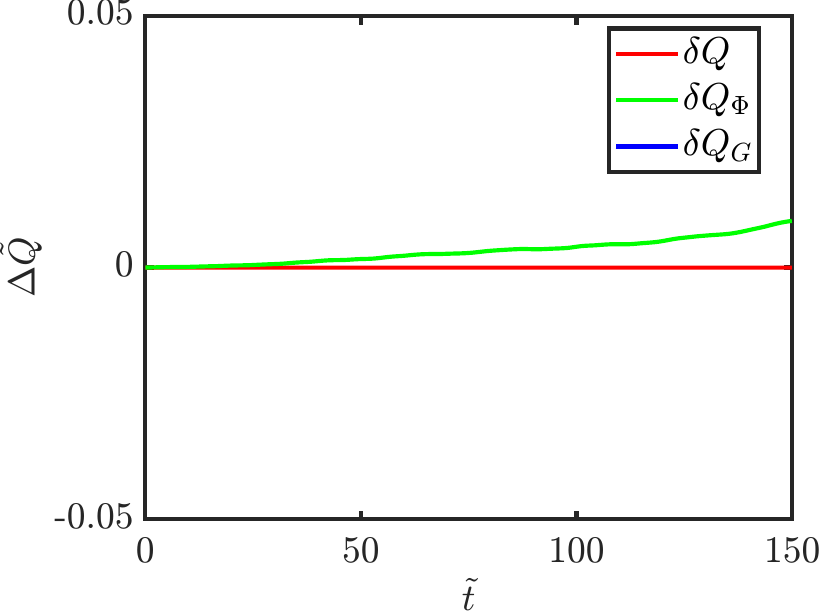}
\caption{Quantum-corrected evolution of the mean fields at a point near the center, their Fourier transformations, the charge evolution and their relative differences from the initial charge in the classical regime ($\omega=0.5h,~g=h=0.1$, with $\alpha=0.01$).\label{fig:10}}
\end{figure}

Figure~\ref{fig:10} shows the corresponding time evolution in the quantum simulation. In the upper-left panel, the mean-field evolution at $(0.25,0.25,0.25)$ in the quantum case (solid line) is very close to the classical result (dashed line). Quantum effects only slightly shift the amplitudes and frequencies. The Fourier spectra, shown in the upper-right panel, are dominated by the frequency $0.5m_{\phi r}$, close to the classical frequency (cf. figure~\ref{fig:7}). The charge evolution shows that the fluctuation sector remains nearly unexcited, so $\tilde{Q}_{\Phi}$ gives an excellent approximation to the total charge. The total charge $\tilde{Q}$ is conserved to high accuracy, and the relative charge deviations remain very small. This confirms that the classical dynamics is a good approximation in this regime. 

\begin{figure}[tbp]
\centering
\includegraphics[width=.44\textwidth]{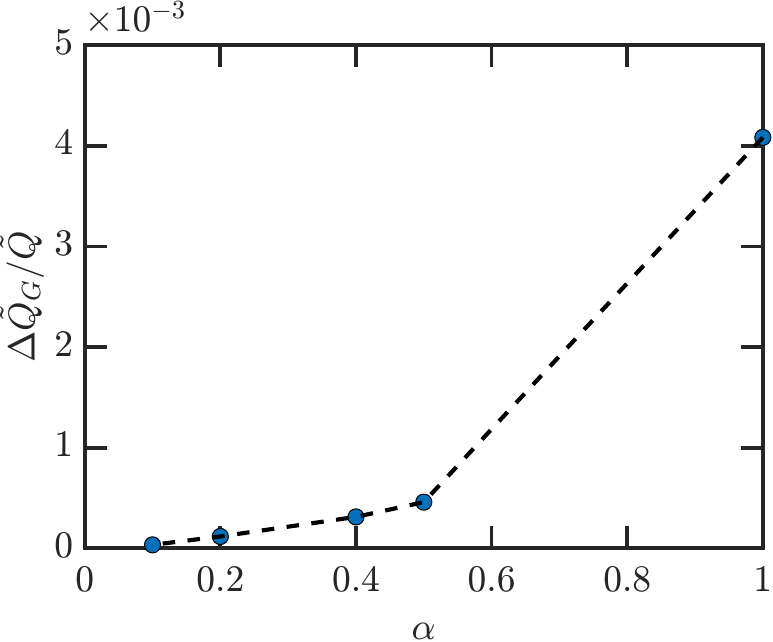}
\caption{Ratio of the fluctuation charge to the total charge for different scaling parameter $\alpha$ (all quantities are time-averaged over the simulation interval from $0-50$). We see a clear transition from the classical regime to the quantum regime as $\alpha$ is increased. \label{fig:14}}
\end{figure}

To illustrate the transition between the classical and quantum regimes, figure~\ref{fig:14} shows the ratio of the quantum fluctuation charge to the total charge for several values of the scaling parameter, with the frequency fixed at $\omega=0.5h$. As $\alpha$ decreases, corresponding to the small-coupling or large-amplitude limit, the effect of quantum fluctuations becomes negligible and the dynamics approaches the classical behavior shown in figure~\ref{fig:10}.

\subsubsection{Quantum Regime}
\label{sec:quantumRegime}

\begin{figure}[tbp]
\centering
\includegraphics[width=.44\textwidth]{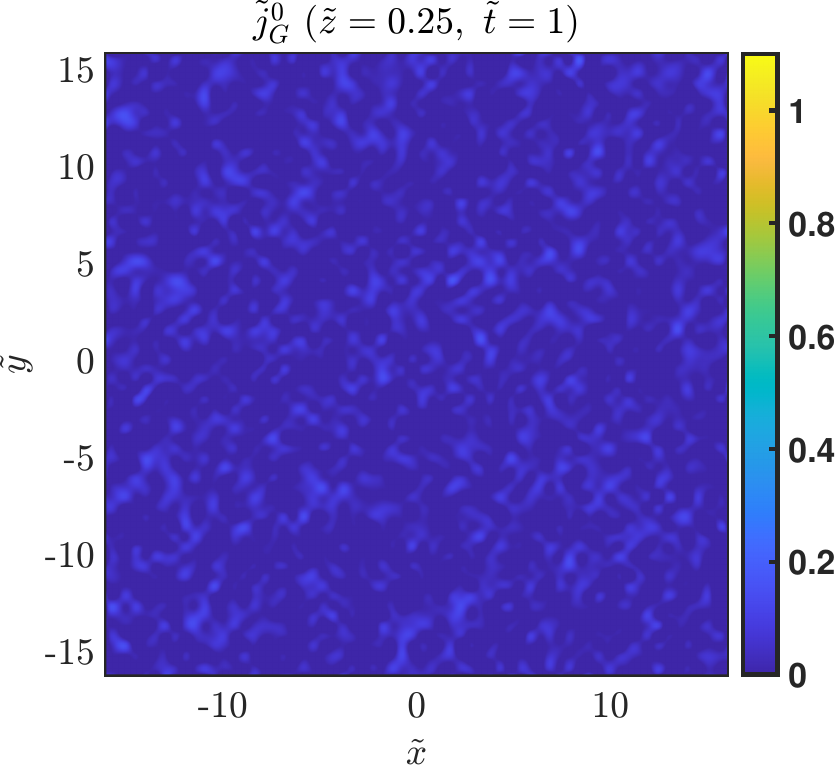}
\hfill
\includegraphics[width=.44\textwidth]{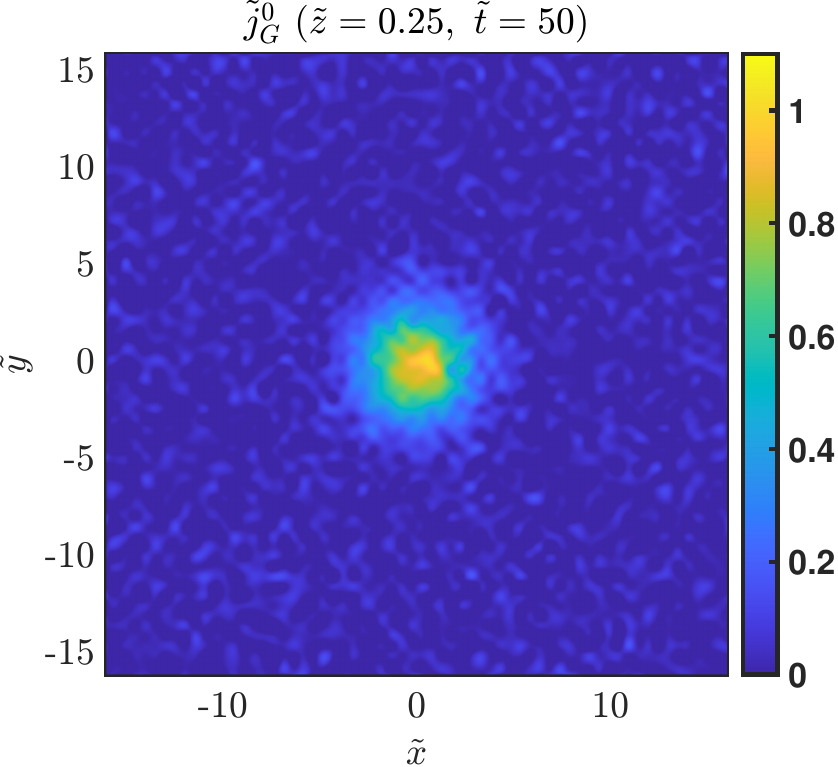}
\caption{Spatial distribution of the charge density $\tilde{j}_G^0$ in a two-dimensional cross section at $\tilde{z}=0.25$ and time $\tilde{t}=1$ (left) and 50 (right) in the quantum regime ($\omega=0.9h,~g=h=0.2$, with $\alpha=0.04$).  \label{fig:13}}
\end{figure}

As the frequency $\omega$ increases, the occupation numbers decrease and quantum fluctuations become more important. The fluctuation contribution to the charge density, $\tilde{j}_G^0$, then becomes significant. Figure~\ref{fig:13} shows its spatial distribution at $\tilde{t}=1$ and $50$. A localized fluctuation charge density develops inside the Q-ball. Although stochastic fluctuations are visible, the distribution appears approximately spherical. To quantify the deviation from perfect spherical symmetry, we compute the $|m|=2$ component of the three-dimensional quadrupole moment using the full data volume 
\begin{equation}
Q_{2}^{\text{(3D)}} = \iiint \tilde{j}_{G}^{0}(\tilde{x},\tilde{y},\tilde{z})_{\tilde{t}=50}\; (\tilde{x}+\mathrm{i}\tilde{y})^{2}\; \mathrm{d}\tilde{x}\,\mathrm{d}\tilde{y}\,\mathrm{d}\tilde{z}.
\end{equation}
This moment is in general complex: its real and imaginary parts measure the two independent $\ell=2$, $|m|=2$ quadrupolar distortions, proportional to $\tilde{x}^{2}-\tilde{y}^{2}$ and $2\tilde{x}\tilde{y}$, respectively. In the following, however, we use only its magnitude $|Q_{2}^{\text{(3D)}}|$, which gives the rotationally invariant amplitude in this $|m|=2$ quadrupole subspace. For strictly spherically symmetric distributions, $Q_{2}^{\text{(3D)}} = 0$. To make the magnitude of $Q_{2}^{\text{(3D)}}$ comparable across different systems, we normalize it by the density-weighted mean squared radius
\begin{equation}
	\bar{Q}_{2} = \frac{|Q_{2}^{\text{(3D)}}|}{\iiint |\tilde{j}_{G}^{0}|_{\tilde{t}=50}\, (\tilde{x}^{2}+\tilde{y}^{2}+\tilde{z}^{2})\; \mathrm{d}\tilde{x}\,\mathrm{d}\tilde{y}\,\mathrm{d}\tilde{z}}.
\end{equation}
This normalized quantity lies in $[0,1]$, with $0$ corresponding to perfect spherical symmetry. For the configuration shown in figure~\ref{fig:13}, we obtain $\bar{Q}_{2} \approx 0.00097$ at $\tilde{t}=50$. The smallness of this value confirms that the distribution is indeed very close to spherical symmetry. The mild quadrupole deformation does not destabilize the configuration; it merely reflects the finite-ensemble stochastic fluctuations.

\begin{figure}[tbp]
\centering
\includegraphics[width=.44\textwidth]{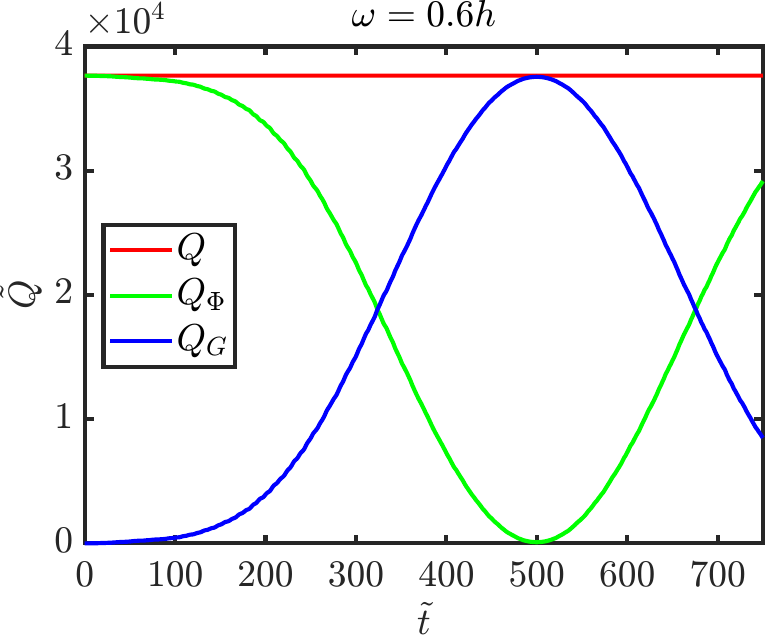}
\hfill
\includegraphics[width=.46\textwidth]{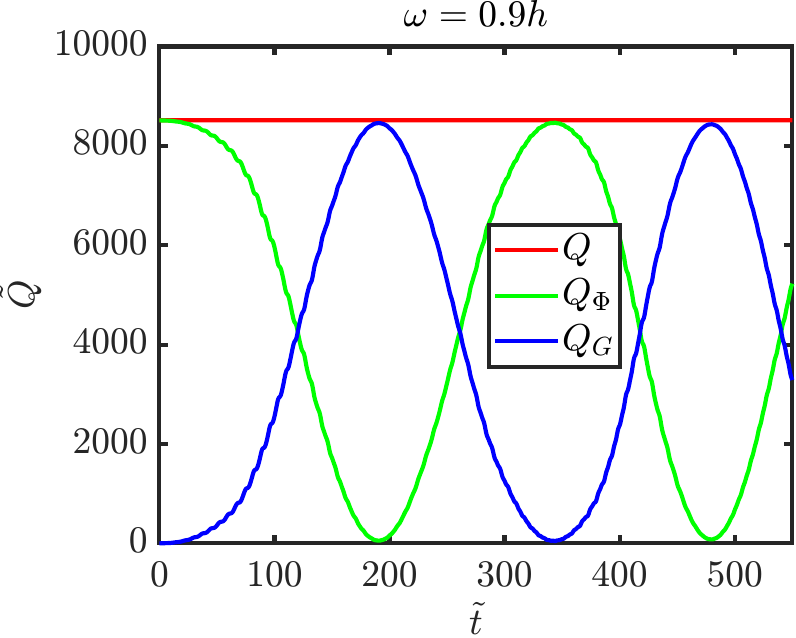}
\caption{Evolution of charges with different frequencies in the quantum regime ($g=h=0.1$, with $\alpha=0.01$). \label{fig:11}}
\end{figure}

\begin{figure}[tbp]
\centering
\includegraphics[width=.44\textwidth]{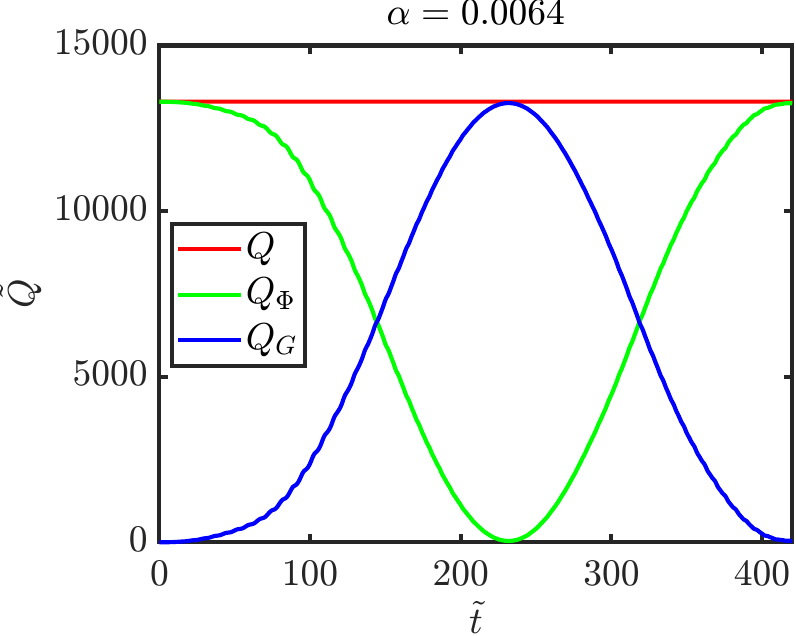}
\hfill
\includegraphics[width=.44\textwidth]{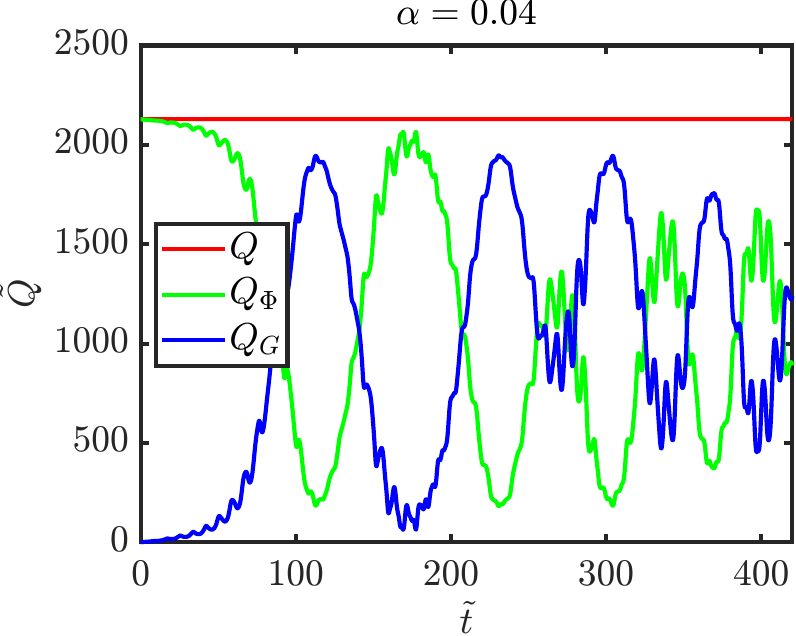}
\hfill
\includegraphics[width=.44\textwidth]{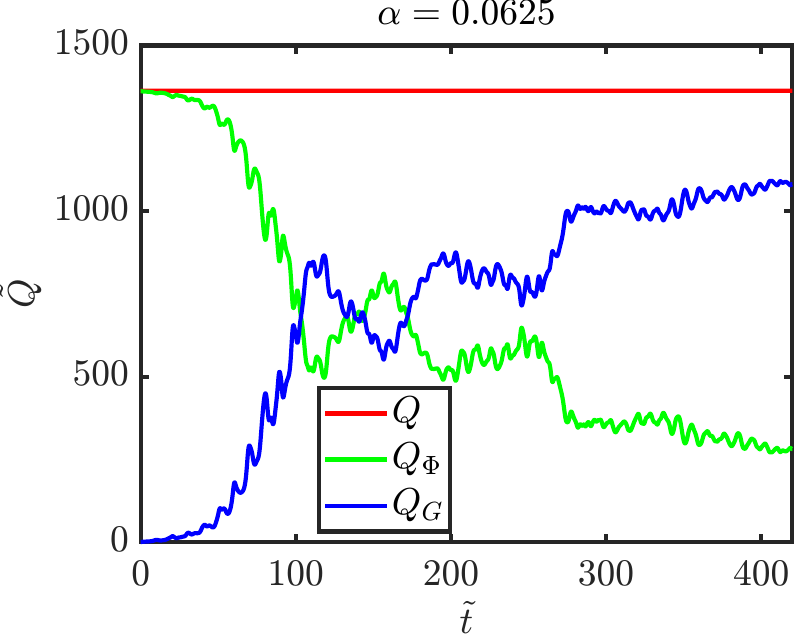}
\hfill
\includegraphics[width=.44\textwidth]{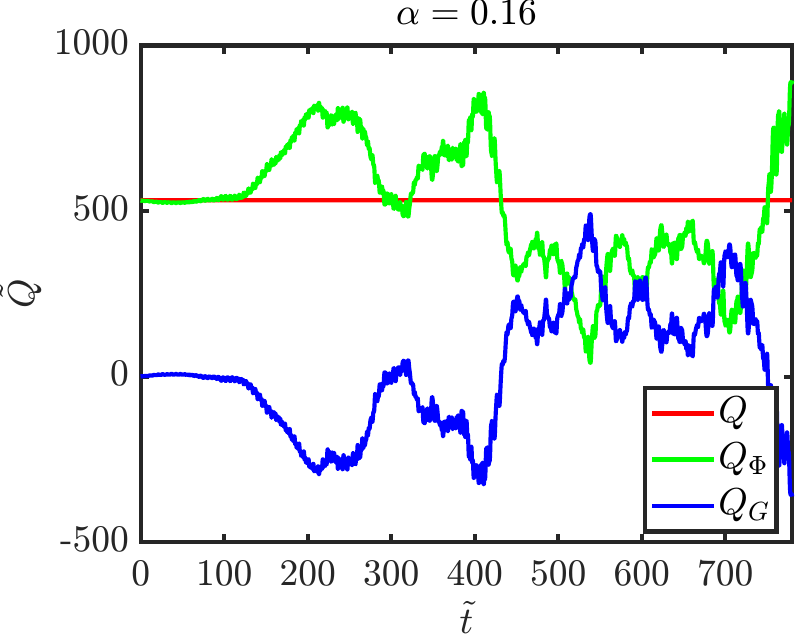}
\caption{Evolution of charges with different scaling parameters ($\omega=0.9h,~g=h=1$).\label{fig:12}}
\end{figure}

The charge evolution shows periodic exchange between the mean fields and the
fluctuation modes, while the total charge $\tilde{Q}$ remains conserved; see
figures~\ref{fig:11} and~\ref{fig:12}.  This exchange can be understood as a
Bogoliubov mixing effect in the rotating Q-ball background.  Expanding
\(\phi=\Phi+\delta\phi\) and \(\chi=\Xi+\delta\chi\), with
\(\Phi=e^{\mathrm{i}\omega t}f(r)/\sqrt2\) and $\Xi=\xi(r)$, the interaction
\(h^2\chi^2|\phi|^2\) contains the quadratic mixing
\begin{equation}
\mathcal L^{(2)}_{\rm mix}
\propto
-2h^2\Xi\,\delta\chi\,
\left(\Phi^*\delta\phi+\Phi\,\delta\phi^*\right).
\end{equation}
This term carries the phases \(e^{\pm \mathrm{i}\omega t}\),
and therefore mixes charged and neutral fluctuation modes, or equivalently
positive- and negative-frequency components in the time-dependent mode
basis.  In the real-field basis this mixing generates the cross correlators
$K_2$ and $K_3$, which act as equal and opposite
sources for the mean-field and fluctuation currents 
\begin{equation}
\partial_\mu j^\mu_\Phi
=
2h^2\Phi_3(\Phi_2K_2-\Phi_1K_3),
\qquad
\partial_\mu j^\mu_G
=
-2h^2\Phi_3(\Phi_2K_2-\Phi_1K_3).
\end{equation}
In a crude background-field estimate, the rotating mean field behaves as
\(\Phi_1+\mathrm{i}\Phi_2\sim e^{\mathrm{i}\omega t}\). The simulation shows that the cross
correlators are dominated by a single rotating component as well, which can be
approximated as  
\(K_2+\mathrm{i}K_3\sim e^{\mathrm{i}\nu t}\). Their
relative rotation gives a beat-like charge transfer
\begin{equation}
\begin{aligned}
\tilde Q_\Phi(t)
&\simeq \bar Q_\Phi
+\Delta Q_-\cos\!\left(|\omega-\nu|t+\delta_-\right)
+\Delta Q_+\cos\!\left(|\omega+\nu|t+\delta_+\right),
\\
\tilde Q_G(t)
&\simeq \bar Q_G
-\Delta Q_-\cos\!\left(|\omega-\nu|t+\delta_-\right)
-\Delta Q_+\cos\!\left(|\omega+\nu|t+\delta_+\right),
\end{aligned}
\end{equation}
with \(\bar Q_\Phi+\bar Q_G=\tilde Q\). The two amplitudes
\(\Delta Q_-\) and \(\Delta Q_+\) depend on the relative weights of the
co-rotating and counter-rotating components and the overall amplitudes are dominated by the low-frequency part. Note this should only be regarded as a rough guide to the observed oscillatory behavior.

Thus \(\tilde{Q}_\Phi\) and \(\tilde{Q}_G\) are not separately conserved and exchange charge with each other, while the total
\(U(1)\) charge is conserved.  At leading Hartree order this transfer is
coherent and not genuinely dissipative, so it appears as an approximately
periodic exchange between the mean-field sector and the fluctuation sector.
Beyond the Hartree approximation, scattering and dissipation may modify this
behavior, which can be studied using higher-order truncations in the 2PI
effective action. However, a detailed study of such effects lies beyond the scope of this work. Here we focus on how the charge exchange depends on the frequency and
scaling parameter, expecting the qualitative picture to remain useful even
when higher-order corrections are included.

In figure~\ref{fig:11}, the scaling parameter is $\alpha=0.01$, equivalently $g=h=0.1$. We show two frequencies, $\omega=0.6h$ and $\omega=0.9h$. In both cases, charge is exchanged periodically between the mean fields and the fluctuation modes, and nearly the entire charge participates in the exchange. The exchange period becomes shorter as the frequency increases and the background excites fluctuation modes more efficiently. Across the parameter ranges we have studied, the system shows a transition from a classical regime, where the Q-ball frequency is small and no appreciable charge exchange occurs, to a quantum regime, where charge exchange appears once the frequency is sufficiently large. 

Next we fix the frequency at $\omega=0.9h$ and vary the scaling parameter $\alpha$. Initially almost all charge resides in the mean fields, up to small fluctuations. Figure~\ref{fig:12} shows that, for very small $\alpha$, the charge exchange is nearly complete. As $\alpha$ increases, the exchange period becomes shorter, but the fraction of charge participating in the exchange decreases: the mean-field and fluctuation sectors exchange charge only partially. For sufficiently large $\alpha$, the charge exchange disappears, indicating that the strong quantum backreaction disrupts the periodic motion of the background fields.

\subsubsection{Stability}
\label{sec:stability}

We now investigate the stability of quantum-corrected Q-balls. Unlike oscillons, which are quasi-stable in both classical and quantum theories \cite{Saffin:2014yka}, $3+1$D Q-balls can be stable or unstable depending on the parameters. This also differs from the $2+1$D case, where Q-balls are always classically stable \cite{Friedberg:1976me}. The classical stability criteria were reviewed in section~\ref{sec:FLS}; here we investigate how genuine quantum effects modify stability within the Hartree approximation \cite{Tranberg:2013cka}. 

\begin{figure}[tbp]
\centering
\includegraphics[width=.42\textwidth]{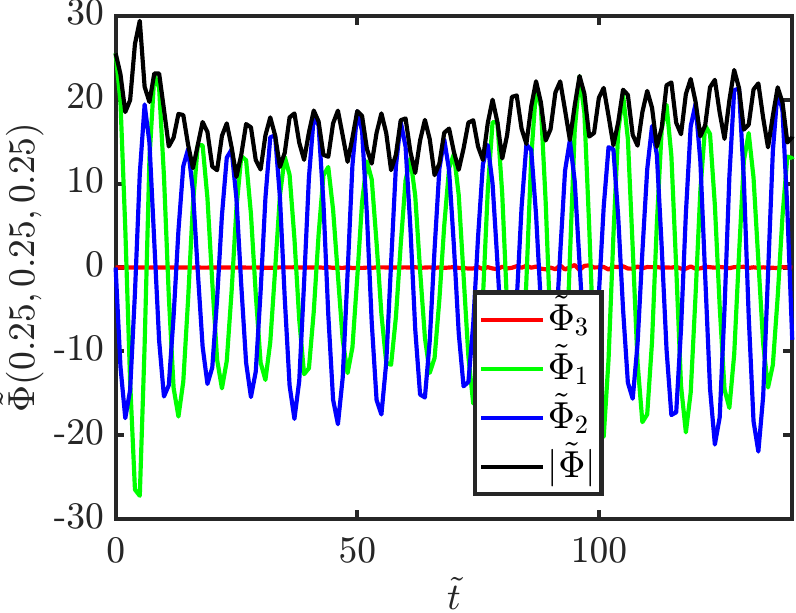}
\hfill
\includegraphics[width=.42\textwidth]{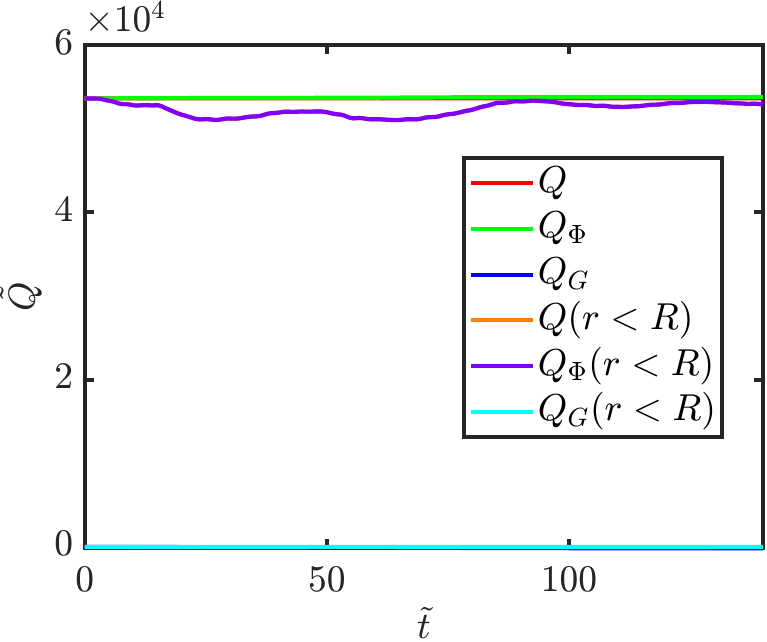}
\hfill
\includegraphics[width=.44\textwidth]{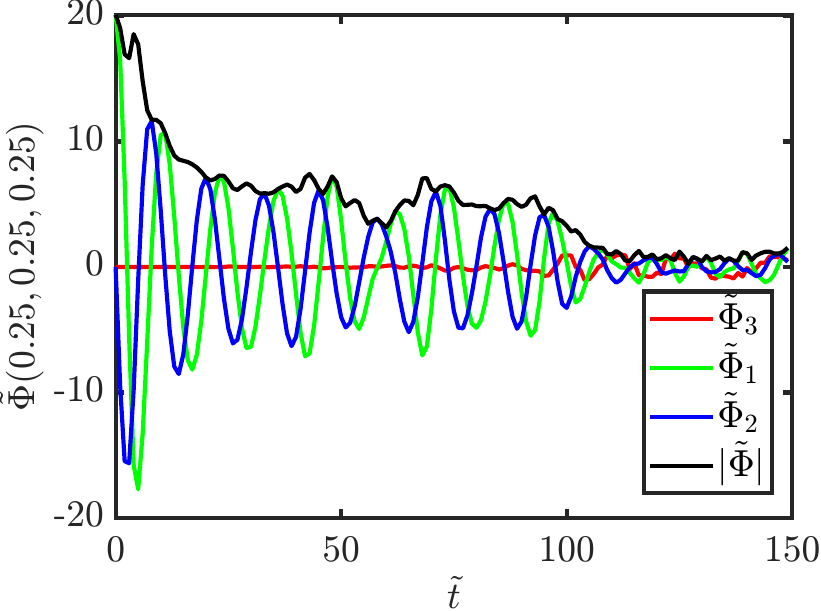}
\hfill
\includegraphics[width=.44\textwidth]{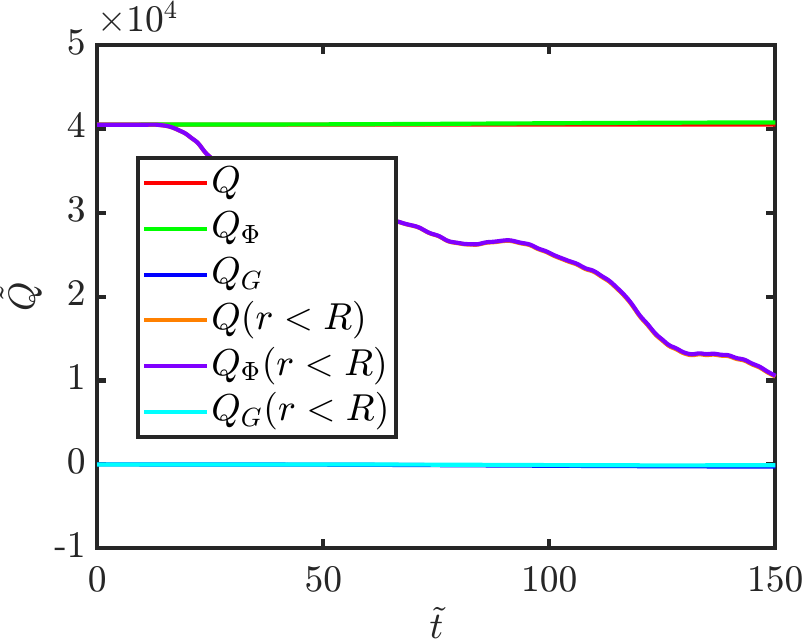}
\hfill
\includegraphics[width=.44\textwidth]{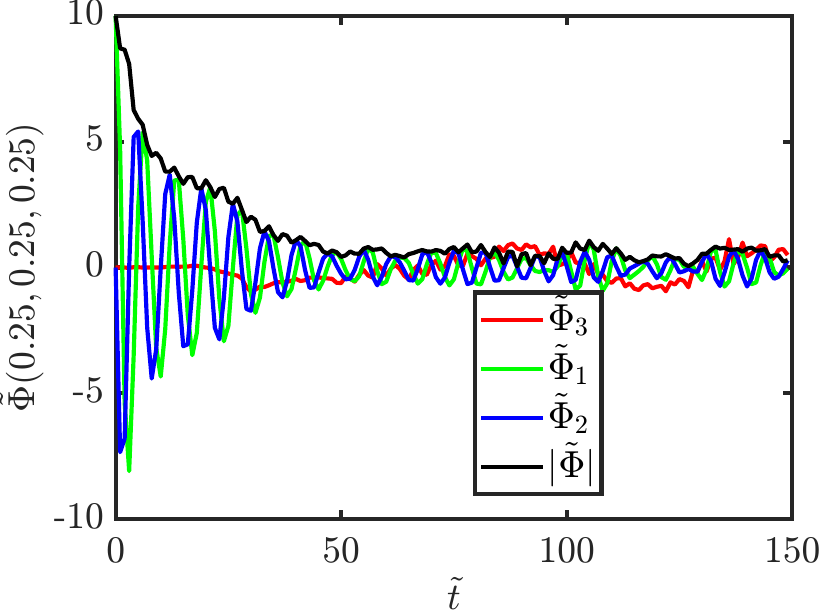}
\hfill
\includegraphics[width=.465\textwidth]{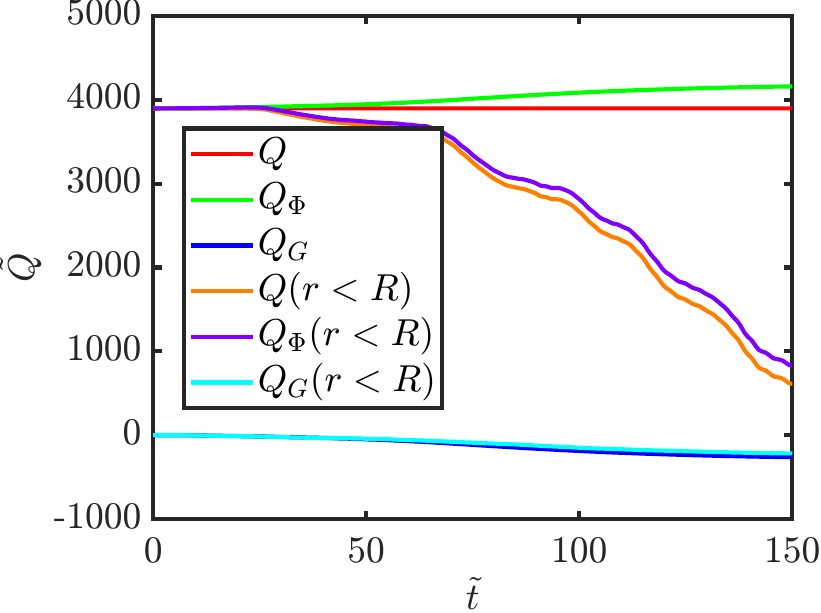}
\caption{(Upper) Evolution of the mean-field and charge of a quantum Q-ball in Case (1) ($\omega/h=0.5,~g=10,h=1$), which is stable under perturbations. (Middle) Evolution of the mean-field and charge in the modified Case (1) with $h=0.6$. The Q-ball is unstable. (Bottom) Evolution of the mean-field and charge in the modified case (1), with $\omega=0.9$. The Q-ball is unstable.}\label{fig:15}
\end{figure}

\begin{figure}[tbp]
\centering
\includegraphics[width=.44\textwidth]{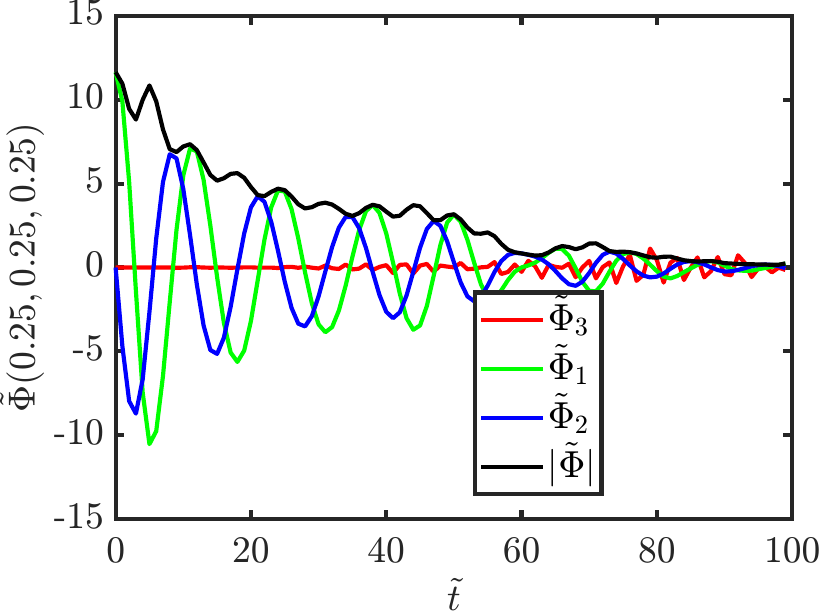}
\hfill
\includegraphics[width=.465\textwidth]{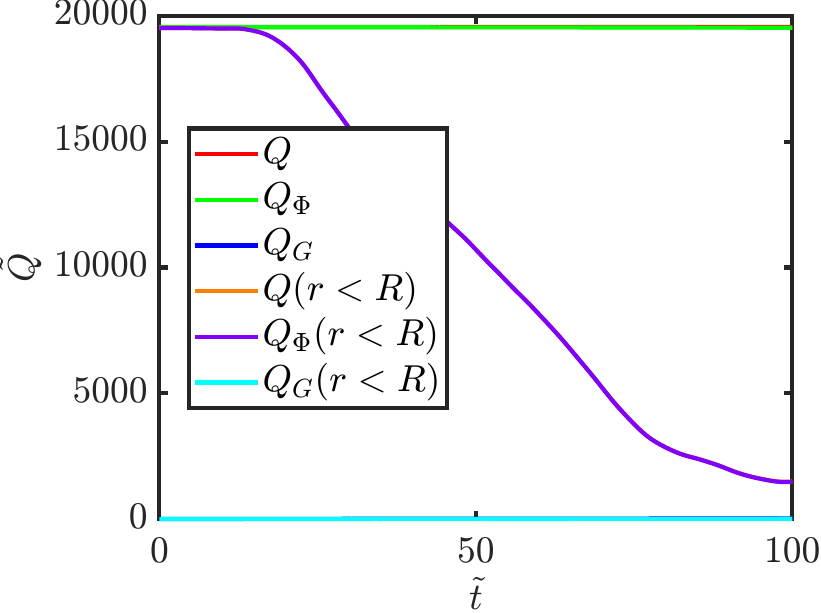}
\hfill
\includegraphics[width=.44\textwidth]{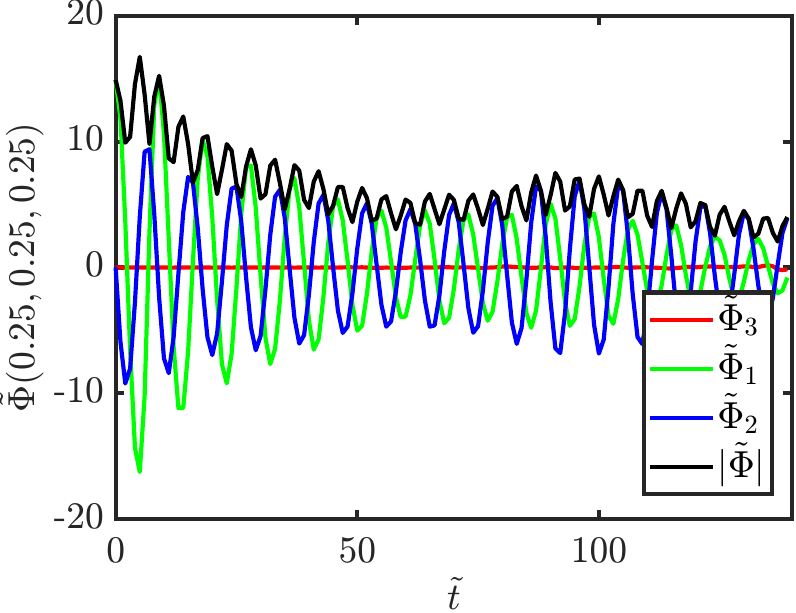}
\hfill
\includegraphics[width=.42\textwidth]{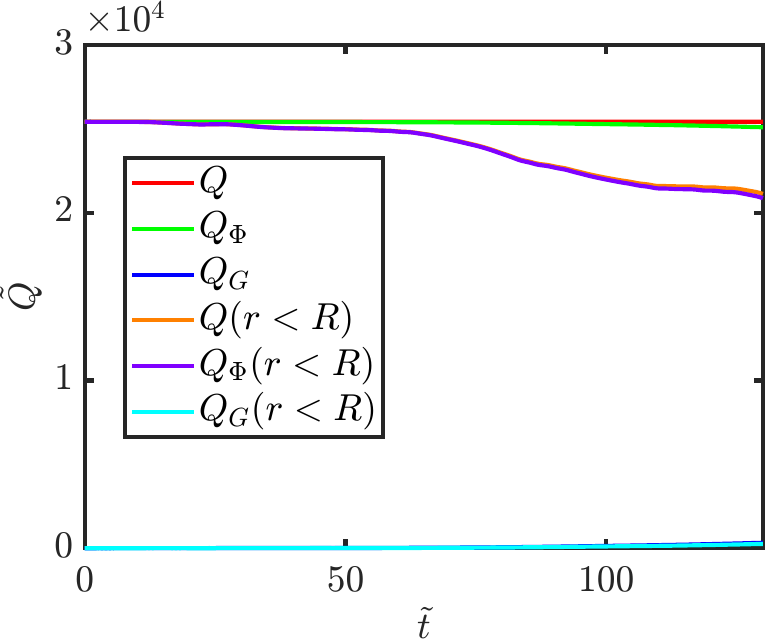}
\caption{(Upper) Evolution of the mean-field and charge of an unstable quantum Q-ball in Case (2) ($\omega/h=0.85,~g=5,h=0.5$). The mean-field oscillates and decays rapidly, and the mean-field charge decays. (Bottom) Evolution of the mean-field and charge in the modified Case (2) with $h=0.8$. The Q-ball is stable under small perturbations.}
\label{fig:16}
\end{figure}

The stability of a Q-ball can be diagnosed from its time evolution. We focus on the mean-field evolution and charge evolution, and find three types of behavior in $3+1$D:
\begin{enumerate}
    \item[(1)] Classically stable and quantum-mechanically stable (e.g. $\omega=0.5,g=10,h=1,\omega/h=0.5$, see the upper row of figure~\ref{fig:15})
    \item[(2)] Classically stable and quantum-mechanically unstable (e.g. $\omega=0.425,g=5,h=0.5, \omega/h=0.85$, see the upper row of figure~\ref{fig:16})
    \item[(3)] Classically unstable and quantum-mechanically unstable (e.g. $\omega=0.475,g=5,h=0.5, \omega/h=0.95$)
\end{enumerate}
An example of Case (1) is shown in the upper row of figure~\ref{fig:15}: the left panel shows the mean-field evolution, and the right panel shows the charge evolution. Both remain stable, apart from small disturbances caused by the initial conditions. An example of Case (2) is shown in the upper row of figure~\ref{fig:16}. In this case, both the mean fields and the charge decay rapidly. Case (3) exhibits a similar decay pattern. This is qualitatively similar to the classical case, but the stability regions are quantitatively different in the parameter space, as shown later. 

We then vary $h$ to study its effect on stability, which we find to be more pronounced than varying $g$ in the parameter ranges considered. Starting from Case (1), we set $h=0.6$, corresponding to $\omega/h = 0.833$; see the middle row of figure~\ref{fig:15}. The quantum evolution becomes unstable, and both the mean-field and the charge decay rapidly after a few periods. Starting from Case (2), we instead set $h=0.8$, corresponding to $\omega/h = 0.531$; see the bottom row of figure~\ref{fig:16}. The quantum evolution is then stable. Finally, increasing the frequency in Case (1) to $\omega=0.9$, or $\omega/h = 0.9$, again makes the quantum evolution unstable; see the bottom row of figure~\ref{fig:15}.

\begin{figure}[tbp]
\centering
\includegraphics[width=.45\textwidth]{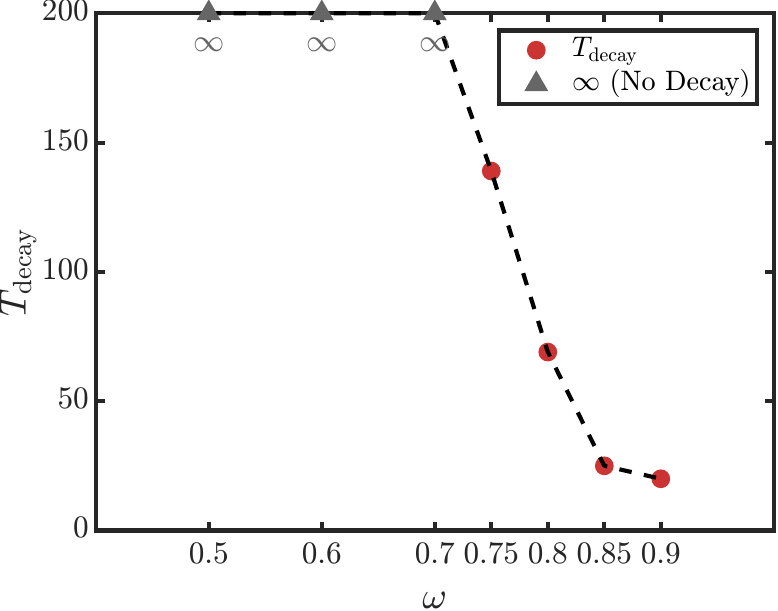}
\caption{Decay time as a function of $\omega$ ($g=10,~h=1$). Quantum Q-balls with frequencies below $\omega\simeq0.75$ do not decay before $\tilde{t}=150$, the maximum duration of our simulations.\label{fig:20}}
\end{figure}

We next examine the dependence on $\omega$ more systematically. We characterize quantum stability using the decay time of $| \tilde{\Phi}|$, defined as the last time at which the field reaches $1/(2e)$ of its initial value within the simulated time interval. Keeping the other parameters fixed and increasing $\omega$ in Case (1), we find the behavior shown in figure~\ref{fig:20}. For low frequencies, the Q-ball does not decay within the simulation time and is therefore stable on the timescales probed. Once $\omega$ exceeds a threshold of about $0.75$, the field begins to decay and the system becomes unstable. Larger $\omega$ leads to a faster decay rate.

The quantum instability region obtained in the inhomogeneous Hartree approximation is larger than the stability estimate from classical solutions. For example, figure~\ref{fig:20} shows quantum instability for $\omega$ above about $0.7$, whereas figure~\ref{fig:4} suggests a classical-limit threshold near $0.86$. This trend is consistent with one-loop calculations \cite{Graham:2001hr}, which show that quantum Q-balls have higher energy than their classical counterparts at fixed charge. 

\begin{figure}[tbp]
\centering
\includegraphics[width=.45\textwidth]{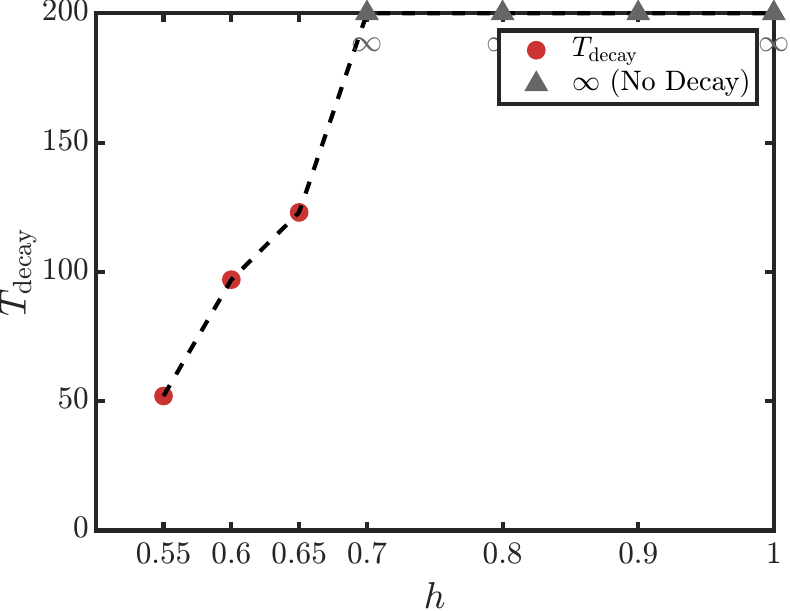}
\caption{Decay time as a function of $h$ ($g=10$, $\omega=0.5$). Quantum Q-balls with $h$ above $h\simeq0.7$ do not decay before $\tilde{t}=150$, the maximum duration of our simulations.\label{fig:22}}
\end{figure}

\begin{figure}[tbp]
\centering
\includegraphics[width=.45\textwidth]{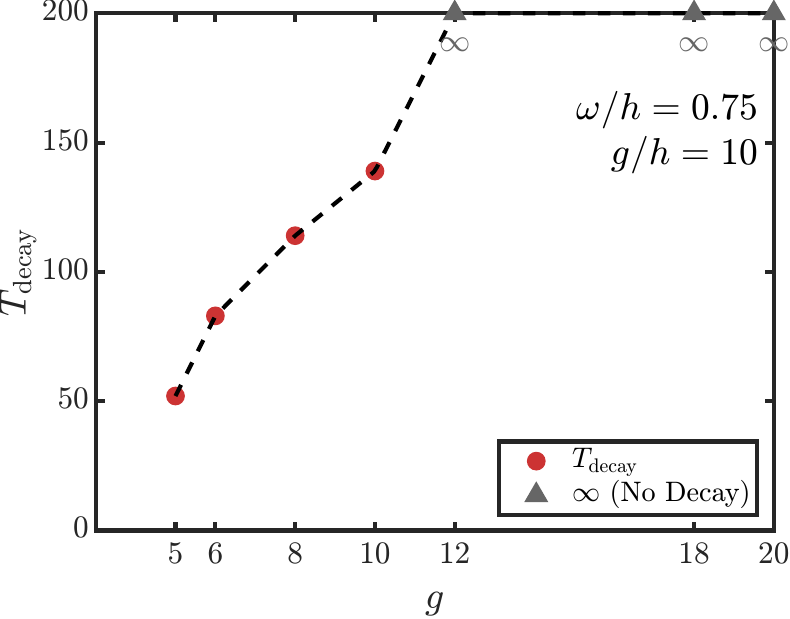}
\caption{Decay time as a function of $g$ ($\omega/h=0.75$, $g/h=10$). Quantum Q-balls with $g$ above $g\simeq12$ do not decay before $\tilde{t}=150$, the maximum duration of our simulations.\label{fig:23}}
\end{figure}

In figure~\ref{fig:22}, we instead vary $h$ in Case (1), keeping the other parameters fixed. The Q-ball remains stable for large $h$, but when $h$ falls below a threshold of about $0.7$, the field begins to decay. Smaller $h$ gives a faster decay rate. Finally, in figure~\ref{fig:23}, we vary $g$ while keeping $\omega/h$ and $g/h$ fixed. The Q-ball remains stable for large $g$, but becomes unstable when $g$ drops below a threshold of about $12$. Again, smaller $g$ gives faster decay. These scans indicate that the frequency ratio $\omega/h$ is a useful indicator
of the onset of instability, while the overall coupling scale also controls the
relative size of Hartree fluctuations and can shift the stability boundary.

\section{Discussion and Outlook}
\label{sec:summary}

In this work, we studied the real-time quantum dynamics of Q-balls in the FLS model. Combining the inhomogeneous Hartree approximation with the stochastic ensemble treatment of the two-point functions, we simulated the coupled evolution of mean fields and quantum fluctuations in $3+1$ dimensions on a lattice. This provides the FLS analogue of earlier Hartree studies of Q-balls in polynomial potentials \cite{Tranberg:2013cka,Xie:2023psz}, and extends them to a renormalizable two-field model. 

The scaling parameter $\alpha$ provides a useful way to organize the approach to the classical limit. In the numerical simulations, small $\alpha$ corresponds to the large-amplitude, highly occupied regime, where the Hartree corrections are relatively suppressed and the evolution remains close to the classical dynamics. As $\alpha$ is increased, quantum backreaction becomes more visible: the fluctuation sector carries a larger fraction of the charge, and the system moves away from the purely classical behavior. This gives a controlled interpolation between the classical regime and the quantum regime.

For fixed model parameters, our results show a clear separation between a classical regime and a quantum regime, depending on the Q-ball frequency. For sufficiently low $\omega/h$, the quantum fluctuation contribution is negligible and the evolution remains close to the classical one. As $\omega/h$ increases, quantum effects become important, and a significant exchange of Noether charge between the mean-field and fluctuation sectors appears. For small $\alpha$ this exchange can be nearly complete, while for larger $\alpha$ it becomes weaker or disappears. This is similar to what is found for $2+1$D Q-balls in polynomial potentials \cite{Xie:2023psz}. 

We also find that quantum fluctuations can affect stability. In particular, besides the region where a Q-ball is both classically and quantum-mechanically stable, and a region where it is unstable, there exists a metastable window in which the Q-ball is classically stable but quantum-mechanically unstable. Although the existence of these regions is found in the classical approximation \cite{Friedberg:1976me}, genuine quantum effects shift the locations of these regions in the parameter space and the stability boundaries should therefore be determined in the quantum-corrected theory. Our numerical scans suggest that the onset of this quantum instability is mainly controlled by $\omega/h$: larger values of $\omega/h$ make the Q-ball more prone to decay. At the same time, the overall coupling scale also affects the stability boundary. In the parameter ranges we have explored, increasing this scale tends to make the configuration longer-lived, while decreasing it makes the decay faster.

There are several natural directions for future work. First, it would be important to extend the present analysis from a single Q-ball to multiple Q-balls. Classical Q-ball systems already exhibit phase-dependent attraction, repulsion, charge transfer and fission \cite{Battye:2000qj}, while charge-swapping Q-balls and recent FLS dipole/chain solutions indicate that multi-soliton dynamics can be considerably richer \cite{Copeland:2014qra,Xie:2021glp,Jaramillo:2024cus}. It would therefore be interesting to study how quantum fluctuations affect scattering, bound-state formation, charge exchange and the stability of multi-Q-ball configurations in the FLS model.
Second, one can go beyond the leading Hartree truncation. In the present approximation, higher-order scattering, dissipation and memory effects are absent. Extending the analysis to higher orders in the $2$PI effective action, or to other controlled nonequilibrium approximation schemes, would help clarify which of the phenomena observed here are quantitatively robust and which are specific to the leading-order truncation \cite{Berges:2000ur,Aarts:2001qa,Berges:2001fi,Berges:2004yj,Aarts:2000wi,Salle:2000hd}. This is especially relevant near the boundary of the metastable region, where small corrections may have a large impact on the lifetime.

\acknowledgments

We would like to thank Guo-Dong Zhang for helpful discussions. SYZ acknowledges support from the National Natural Science
Foundation of China under grant No.~12475074 and No.~12247103. QXX acknowledges support from CSC (File No.~202406340173).

~

\appendix

\noindent{\bf APPENDIX}

\section{Hartree Approximation From 2PI Effective Action}
\label{app:2pi}

In this appendix, we show that the dynamics obtained from the leading local truncation of the 2PI effective action is equivalent to the Hartree approximation \cite{Calzetta:2008iqa}. In this truncation, connected correlators of order higher than two are set to zero, leaving only the mean fields and connected two-point correlators. 

We denote the three real quantum fields by
\[
\psi_i=(\phi_1,\phi_2,\chi),\qquad i=1,2,3.
\]
Their mean fields are
\[
\Phi_i(x)=\langle \psi_i(x)\rangle ,
\]
and the fluctuation fields are
\[
\delta\psi_i(x)=\psi_i(x)-\Phi_i(x).
\] 
We first extend the path integral by adding nonlocal sources. After introducing a bilocal source, the path integral can be written as 
\begin{equation}
    Z\left[ J,R \right] =\int{\mathcal{D} \psi \,\,\exp \left[ \mathrm{i}\left( S\left[ \psi \right] +J_i\psi ^i+\frac{1}{2}\psi ^iR_{ij}\psi ^j \right) \right]},
\end{equation}
where $J_i$ is the local source, $R_{ij}$ is the bilocal source and the path integral is defined on the closed time path.

The connected generating functional is defined by
\begin{equation}
    W\left[ J,R \right] =-\mathrm{i}\ln Z\left[ J,R \right]. 
\end{equation}
The mean fields $\Phi^i$ and the connected two-point correlators $\mathcal G^{ij}$ are\footnote{In this appendix we use the physical contour-ordered connected correlator
\[
\mathcal G_{ij}(x,y)
=
\langle \mathrm{T}_\mathcal{C}\delta\psi_i(x)\delta\psi_j(y)\rangle,
\]
with \(\mathrm{T}_{\mathcal C}\) denoting contour ordering along the closed time path \(\mathcal C\).
We do not absorb any extra factor of \(\mathrm{i}\) into the definition of \(\mathcal G_{ij}\).} 
\begin{equation}
    \frac{\delta W}{\delta J_i} = \Phi^i ,\quad \frac{\delta W}{\delta R_{ij}} = \frac{1}{2}(\mathcal G^{ij}+\Phi^i\Phi^j).
\end{equation}
The 2PI effective action is the Legendre transform of the generating functional with respect to the local and bilocal sources $J,R$
\begin{equation}
    \Gamma \left[ \Phi ,\mathcal G \right] =W\left[ J,R \right] -J_i\frac{\delta W}{\delta J_i}-R_{ij}\frac{\delta W}{\delta R_{ij}}=W\left[ J,R \right] -J_i\Phi ^i-\frac{1}{2}R_{ij}\left( \Phi ^i\Phi ^j+\mathcal G^{ij} \right) .
\end{equation}
This gives the stationary conditions
\begin{align}
    \frac{\delta \Gamma \left[ \Phi ,\mathcal G \right]}{\delta \Phi ^i}&=-J_i-R_{ij}\Phi ^j,
\\
    \frac{\delta \Gamma \left[ \Phi ,\mathcal G \right]}{\delta \mathcal G^{ij}}&=-\frac{1}{2}R_{ij}.
\end{align}
At vanishing sources, the equations of motion of physical mean fields and propagators are therefore determined 
\begin{equation}
    \frac{\delta\Gamma[\Phi,\mathcal G]}{\delta\Phi^i}=0,
    \qquad
    \frac{\delta\Gamma[\Phi,\mathcal G]}{\delta\mathcal G^{ij}}=0.
\end{equation}

The 2PI effective action can be written as
\begin{equation}
    \Gamma \left[ \Phi ,\mathcal G \right] =S\left[ \Phi \right] +\frac{\mathrm{i}}{2}~\mathrm{Tr}\left[ \ln \mathcal G^{-1}+\left( \mathcal G_{0}^{-1}-\mathcal G^{-1} \right) \mathcal G \right] -\Phi_{\rm 2PI} \left[ \Phi ,\mathcal G \right] ,
\end{equation}
where \(\Phi_{\rm 2PI}\) denotes the sum of vacuum two-particle-irreducible skeleton diagrams with full propagator lines \(\mathcal G\), which can also be obtained by shifting the classical action above background fields. It should not be confused with the mean fields \(\Phi_i\). The classical inverse propagator is defined by
\begin{equation}
    \mathrm{i}\mathcal G^{-1}_{0,ij}(x,y;\Phi)
    =
    \left.
    \frac{\delta^2 S[\psi]}
    {\delta\psi_i(x)\delta\psi_j(y)}
    \right|_{\psi=\Phi}.
\end{equation}

In our study, we use the Hartree, or local bubble, truncation of the 2PI effective action. Nonlocal sunset diagrams generated by cubic vertices in the shifted action are not included. Consequently the self-energy is local, and the equations close on the mean fields and connected two-point functions.

For the FLS model, the action is
\begin{equation}
    S[\psi]=\int_{\mathcal C} \mathrm{d}^4x
\left[
\frac12\sum_{i=1}^3 \partial_\mu\psi_i\partial^\mu\psi_i
-U(\psi)
\right],
\end{equation}
with
\begin{equation}
    U(\psi)
=
\frac12 h^2\psi_3^2(\psi_1^2+\psi_2^2)
+\frac12 w_1^2(\psi_1^2+\psi_2^2)
+\frac12 w_2^2\psi_3^2
+\frac18 g^2(\psi_3^2-\chi_v^2)^2 .
\end{equation}
We use the closed-time-path formalism, integrating along the real-time Schwinger-Keldysh contour $\mathcal{C} $ from an initial time $t_0$ to time $t$ along $\mathcal{C}_+ $ and then back to $t_0$ along $\mathcal{C}_-$. 

The propagators can be written in matrix form as
\begin{equation}
    \mathcal{G} \left( x,y \right) =\left( \begin{matrix}
	G_1\left( x,y \right)&		K_1\left( x,y \right)&		K_2\left( x,y \right)\\
	\bar{K}_1\left( x,y \right)&		G_2\left( x,y \right)&		K_3\left( x,y \right)\\
	\bar{K}_2\left( x,y \right)&		\bar{K}_3\left( x,y \right)&		G_3\left( x,y \right)\\
\end{matrix} \right) .
\end{equation}
Here and in the following equations before the Wightman decomposition~\eqref{eq:wightman_decomp},
\(G_i,K_i,\bar K_i\) denote the contour-ordered components of
\(\mathcal G\). They should not yet be identified with the real-time
Wightman two-point functions used in section~\ref{sec:hatree}.
More explicitly, they are defined by
\begin{align} 
\label{eq:timeorder_cor1}
G_1\left( x,y \right) &=\left< \mathrm{T}_{\mathcal{C}}\phi _1\left( x \right) \phi _1\left( y \right) \right> -\left< \phi _1\left( x \right) \right> \left< \phi _1\left( y \right) \right>,  
\\
G_2\left( x,y \right) &=\left< \mathrm{T}_{\mathcal{C}}\phi _2\left( x \right) \phi _2\left( y \right) \right> -\left< \phi _2\left( x \right) \right> \left< \phi _2\left( y \right) \right> ,
\\
G_3\left( x,y \right) &=\left< \mathrm{T}_{\mathcal{C}}\chi \left( x \right) \chi \left( y \right) \right> -\left< \chi \left( x \right) \right> \left< \chi \left( y \right) \right> ,
\\
K_1\left( x,y \right) &=\left< \mathrm{T}_{\mathcal{C}}\phi _1\left( x \right) \phi _2\left( y \right) \right> -\left< \phi _1\left( x \right) \right> \left< \phi _2\left( y \right) \right> ,
\\
\bar{K}_1\left( x,y \right) &=\left< \mathrm{T}_{\mathcal{C}}\phi _2\left( x \right) \phi _1\left( y \right) \right> -\left< \phi _2\left( x \right) \right> \left< \phi _1\left( y \right) \right> ,
\\
K_2\left( x,y \right) &=\left< \mathrm{T}_{\mathcal{C}}\phi _1\left( x \right) \chi \left( y \right) \right> -\left< \phi _1\left( x \right) \right> \left< \chi \left( y \right) \right> ,
\\
\bar{K}_2\left( x,y \right) &=\left< \mathrm{T}_{\mathcal{C}}\chi \left( x \right) \phi _1\left( y \right) \right> -\left< \chi \left( x \right) \right> \left< \phi _1\left( y \right) \right>, 
\\
K_3\left( x,y \right) &=\left< \mathrm{T}_{\mathcal{C}}\phi _2\left( x \right) \chi \left( y \right) \right> -\left< \phi _2\left( x \right) \right> \left< \chi \left( y \right) \right> ,
\\
\label{eq:timeorder_cor2}
\bar{K}_3\left( x,y \right) &=\left< \mathrm{T}_{\mathcal{C}}\chi \left( x \right) \phi _2\left( y \right) \right> -\left< \chi \left( x \right) \right> \left< \phi _2\left( y \right) \right> .
\end{align}
The inverse propagator of the classical theory is
\begin{equation}
\begin{aligned}
\mathrm{i}\mathcal{G}_{0}^{-1}(x,y)&= \begin{pmatrix}
-\Box - h^2 \Phi_3^2 - w_1^2 & 0 & -2h^2 \Phi_1 \Phi_3 \\
0 & -\Box - h^2 \Phi_3^2 - w_1^2 & -2h^2 \Phi_2 \Phi_3 \\
-2h^2 \Phi_1 \Phi_3 & -2h^2 \Phi_2 \Phi_3 & -\Box - h^2 (\Phi_1^2 + \Phi_2^2) - \frac{1}{2} g^2 (3\Phi_3^2 - \chi_v^2) - w_2^2
\end{pmatrix} \\
&\quad \times \delta_\mathcal{C}^{(4)}(x-y).
\end{aligned}
\end{equation}
The contour delta function is defined by
\[
\int_\mathcal{C} \mathrm{d}^4 z\,\delta_\mathcal{C}^{(4)}(x-z)F(z)=F(x).
\]
where $F(x)$ is an arbitrary function. Equivalently, if \(x\) and \(y\) lie on the branches \(a,b=\pm\) of the
Schwinger-Keldysh contour, then
\[
\delta_\mathcal{C}^{(4)}(x_a-y_b)
=
\eta_a\,\delta_{ab}\,
\delta(x^0-y^0)\delta^{(3)}(\mathbf x-\mathbf y),
\qquad
\eta_+=1,\quad \eta_-=-1 .
\]
The sign \(\eta_a\) reflects the orientation of the corresponding branch of the closed time path.

Keeping only the leading-order contribution to $\Phi_{\rm 2PI}$, namely the double-bubble diagrams, gives the truncated functional
\begin{equation}
    \Phi_{\rm 2PI}=\frac{h^2}{2}\int_{\mathcal{C}}{\left[ G_3\left( G_1+G_2 \right) +2(K_{2}\bar K_{2}+K_{3}\bar K_{3}) \right]}\mathrm{d}^4x+\frac{3g^2}{8}\int_{\mathcal{C}}{G_{3}^{2}}\mathrm{d}^4x.
\end{equation}
The corresponding truncated 2PI effective action is
\begin{equation}
        \Gamma_{\rm trunc} =S +\frac{\mathrm{i}}{2}~\mathrm{Tr}\left[ \ln \mathcal{G} ^{-1}+\left( \mathcal{G} _{0}^{-1}-\mathcal{G} ^{-1} \right) \mathcal{G}  \right] -\Phi_{\rm 2PI},
\end{equation}
and the equations of motion follow from the stationary conditions
\begin{equation}
    \frac{\delta \Gamma _{\rm trunc}}{\delta \Phi _i}=0, \quad \frac{\delta \Gamma _{\rm trunc}}{\delta \mathcal{G}_{ij}}=0.
\end{equation}

The first condition gives Eqs.~\eqref{eq:50}-\eqref{eq:52}. The second condition ${\delta \Gamma _{\rm trunc}}/{\delta \mathcal{G}}=0$ gives the Dyson equation
\begin{equation}
    \mathrm{i}\mathcal{G} ^{-1}=\mathrm{i}\mathcal{G} _{0}^{-1}-\Sigma, 
\end{equation}
where the self-energy is
\begin{equation}
\Sigma_{ij}(x,y)=2\frac{\delta \Phi _{\rm 2PI}}{\delta \mathcal{G}_{ij}(x,y)}=\left( \begin{matrix}
	h^2G_3&		0&		2h^2K_2\\
	0&		h^2G_3&		2h^2K_3\\
	2h^2\bar K_{2}&		2h^2\bar K_{3}&		h^2\left( G_1+G_2 \right) +\frac{3}{2}g^2G_3\\
\end{matrix} \right) \delta_\mathcal C^{(4)}(x-y),
\end{equation}
and the matrix $\mathrm{i}\mathcal{G}_{ij}^{-1}(x,y)$ is
\begin{equation}
\begin{aligned}
&
\begin{pmatrix}
-\Box -h^2\Phi _{3}^{2}-w_{1}^{2}-h^2G_3 & 0 & -2h^2\Phi _1\Phi _3-2h^2K_2 \\
0 & -\Box -h^2\Phi _{3}^{2}-w_{1}^{2}-h^2G_3 & -2h^2\Phi _2\Phi _3-2h^2K_3 \\
-2h^2\Phi _1\Phi _3-2h^2\bar K_{2} & -2h^2\Phi _2\Phi _3-2h^2\bar K_{3} & 
\!\begin{aligned}
-\Box &-h^2\left( \Phi _{1}^{2}+\Phi _{2}^{2} \right)  -\frac{1}{2}g^2\left( 3\Phi _{3}^{2}-\chi _{v}^{2} \right) \\
 & -w_{2}^{2}-h^2\left( G_1+G_2 \right) -\frac{3}{2}g^2G_3 
\end{aligned}
\end{pmatrix}\\
&\times\delta_\mathcal{C}^{(4)}(x-y).
\end{aligned}
\end{equation}
Finally, we have the contour-ordered propagator equations, which contain contact terms proportional to
\(\delta_\mathcal{C}^{(4)}(x-y)\).

To obtain the real-time equations in section \ref{sec:hatree}, which are suitable for numerical evolution, we decompose the contour-ordered propagators into Wightman functions. On the closed time path, the two-point functions can be expressed as
\begin{equation}
\label{eq:wightman_decomp}
    \mathcal{G}(x,y) = \Theta_{\mathcal{C}}(x_0-y_0)\, \mathcal{G}^{>}(x,y) + \Theta_{\mathcal{C}}(y_0-x_0)\, \mathcal{G}^{<}(x,y),
\end{equation}
where \[
\mathcal G^{>}_{ij}(x,y)
=
\langle \delta\psi_i(x)\delta\psi_j(y)\rangle,
\qquad
\mathcal G^{<}_{ij}(x,y)
=
\langle \delta\psi_j(y)\delta\psi_i(x)\rangle ,
\] are the positive and negative frequency Wightman functions. Notice that the objects
defined in Eqs.~\eqref{eq:timeorder_cor1}-\eqref{eq:timeorder_cor2} are contour-ordered propagators and are
therefore not identical to the real-time two-point functions introduced
in section~\ref{sec:hatree}. The latter are obtained by projecting the
contour propagator onto its positive frequency component, e.g.
\(G_1=\mathcal G^{>}_{11}\), \(K_1=\mathcal G^{>}_{12}\),
\(K_2=\mathcal G^{>}_{13}\), and similarly for the other components. For these Wightman components, the contour delta functions do not contribute, and one obtains the real-time evolution Eqs. \eqref{eq:57}-\eqref{eq:65}.

The derivation of the Hartree equations does not require specifying a particular initial state. In the numerical implementation of the main text, however, we choose Gaussian initial fluctuations. This allows the Wightman functions to be represented in terms of mode functions. Substituting this expansion into the above equations and using mode orthogonality yields
\begin{equation}
    \left[ \partial _{x}^{2}+\bar{M}_{1}^{2}\left( x \right) \right] f_{\mathbf{k}}^{1}\left( x \right) +\bar{M}_{2}^{2}\left( x \right) f_{\mathbf{k}}^{3}\left( x \right) =0,
\end{equation}
\begin{equation}
\left[ \partial _{x}^{2}+\bar{M}_{1}^{2}\left( x \right) \right] f_{\mathbf{k}}^{2}\left( x \right) +\bar{M}_{3}^{2}\left( x \right) f_{\mathbf{k}}^{3}\left( x \right) =0,
\end{equation}
\begin{equation}
\left[ \partial _{x}^{2}+\bar{M}_{4}^{2}\left( x \right) \right] f_{\mathbf{k}}^{3}\left( x \right) +\bar{M}_2\left( x \right) ^2f_{\mathbf{k}}^{1}\left( x \right) +\bar{M}_{3}^{2}\left( x \right) f_{\mathbf{k}}^{2}\left( x \right) =0.
\end{equation}
These coincide with Eqs. \eqref{eq:mode1}-\eqref{eq:mode3}.

Thus, the leading-order 2PI truncation reproduces the same equations as the inhomogeneous Hartree approximation used in the main text.

\bibliographystyle{JHEP}
\bibliography{biblio}

\end{document}